\documentclass[useAMS,usenatbib,usedcolumn]{mn2e}

\def\HI{\ifmmode{\rm HI}\else{H\/{\sc i}}\fi}

\def\sun{\hbox{\scriptsize $\odot$}}
\def\lsun{\ifmmode{{\mathrm L}_{\odot}}\else{L$_{\odot}$}\fi}

\def\msun{\ifmmode{{\mathrm M}_{\odot}}\else{M$_{\odot}$}\fi} 
\def\msunpc2{\ifmmode{{\mathrm M}_{\odot} \, {\mathrm{pc}}^{-2}}\else{M$_{\odot} \, {\mathrm {pc}}^{-2}$}\fi}

\def\kms{\ifmmode{{\mathrm{km \, s^{-1}}}}\else{${\mathrm{km \, s^{-1}}}$}\fi}

\def\la{\mathrel{\mathchoice {\vcenter{\offinterlineskip\halign{\hfil
$\displaystyle##$\hfil\cr<\cr\sim\cr}}}
{\vcenter{\offinterlineskip\halign{\hfil$\textstyle##$\hfil\cr
<\cr\sim\cr}}}
{\vcenter{\offinterlineskip\halign{\hfil$\scriptstyle##$\hfil\cr
<\cr\sim\cr}}}
{\vcenter{\offinterlineskip\halign{\hfil$\scriptscriptstyle##$\hfil\cr
<\cr\sim\cr}}}}}
\def\ga{\mathrel{\mathchoice {\vcenter{\offinterlineskip\halign{\hfil
$\displaystyle##$\hfil\cr>\cr\sim\cr}}}
{\vcenter{\offinterlineskip\halign{\hfil$\textstyle##$\hfil\cr
>\cr\sim\cr}}}
{\vcenter{\offinterlineskip\halign{\hfil$\scriptstyle##$\hfil\cr
>\cr\sim\cr}}}
{\vcenter{\offinterlineskip\halign{\hfil$\scriptscriptstyle##$\hfil\cr
>\cr\sim\cr}}}}}

\def\aj{AJ}
\def\araa{ARA\&A}
\def\apj{ApJ}
\def\apjl{ApJ}
\def\apjs{ApJS}
\def\aap{A\&A}
\def\mnras{MNRAS}
\def\pasp{PASP}
\def\pasj{PASJ}
\def\nat{Nature}

\usepackage[figuresright]{rotating}
\usepackage{lscape}
\usepackage{graphics}
\usepackage{epsfig}
\usepackage{multirow}
\usepackage{bigdelim}
\usepackage{bigstrut}
\usepackage{amsmath}
\usepackage{amssymb}
\usepackage{hyperref}
\usepackage{epstopdf}

\defcitealias{Jaffe2016}{J16}

\title[BUDHIES III]{BUDHIES III: 
The fate of \HI\ and the quenching of galaxies in evolving environments\\
}
\author[Y.~Jaff\'e et al.] {Yara L. Jaff\'e$^{1,2}$\thanks{E-mail: yjaffe@eso.org}, 
Marc A.~W. Verheijen$^3$, 
Chris P. Haines$^{4,5}$, 
Hyein Yoon$^6$, 
\and Ryan Cybulski$^7$, 
Mar\'{i}a Montero-Casta\~no$^8$, 
Rory  Smith$^6$, 
Aeree Chung$^6$,  
\and Boris Z. Deshev$^{3,9,10}$, 
Ximena Fern\'{a}ndez$^{11}$, 
Jacqueline van Gorkom$^{12}$, 
\and Bianca M. Poggianti$^{13}$,
Min S. Yun$^7$, 
Alexis Finoguenov$^{14}$, 
Graham P. Smith$^{15}$, 
\and Nobuhiro Okabe$^{16,17}$ \\
   $^1$European Southern Observatory, Alonso de Cordova 3107, Vitacura, Casilla 19001, Santiago de Chile, Chile \\
   $^2$Department of Astronomy, Universidad de Concepci\'on, Casilla 160-C, Concepci\'on, Chile\\
   $^3$Kapteyn Astronomical Institute, University of Groningen, Landleven 12, NL-9747 AD-Groningen, The Netherlands\\
   $^4$Departamento de Astronom\'ia, Universidad de Chile, Casilla 36-D, Correo Central, Santiago, Chile\\  
   $^5$INAF-Osservatorio Astronomico di Brera, via Brera 28, I-20122 Milano, Italy\\
   $^6$Department of Astronomy, Yonsei University, 50 Yonsei-ro, Seodaemun-gu, Seoul 120-749, Korea\\
   $^7$Department of Astronomy, University of Massachusetts, 710 North Pleasant Street, Amherst, MA 01003, USA \\
   $^8$Dunlap Institute for Astronomy and Astrophysics, University of Toronto, 50 St George Street, Toronto, ON M5S 3H4, Canada \\
   $^9$Tartu Observatory, T\~oravere, 61602, Estonia\\
   $^{10}$Institute of Physics, University of Tartu, Ravila 14c, 50411, Estonia \\
   $^{11}$Department of Physics and Astronomy, Rutgers, The State University of New Jersey, Piscataway, NJ 08854-8019, USA \\
   $^{12}$Department of Astronomy, Columbia University, Mail Code 5246, 550 W 120th Street, New York, NY 10027, USA\\
   $^{13}$INAF-Osservatorio Astronomico di Padova, vicolo dell' Osservatorio 5, I-35122 Padova, Italy\\
   $^{14}$Department of Physics, University of Helsinki, Gustaf H\"allstr\"omin katu 2a, FI-0014 Helsinki, Finland\\
   $^{15}$School of Physics and Astronomy, University of Birmingham, Edgbaston, Birmingham, B15 2TT, UK\\
   $^{16}$Department of Physical Science, Hiroshima University, 1-3-1 Kagamiyama, Higashi-Hiroshima, Hiroshima 739-8526, Japan\\ 
   $^{17}$Hiroshima Astrophysical Science Center, Hiroshima University, Higashi-Hiroshima, Kagamiyama 1-3-1, 739-8526, Japan
   }
\begin{document}

\maketitle

\begin{abstract} 
In a hierarchical Universe clusters grow via the accretion of galaxies from the field, groups and even other clusters. 
As this happens, galaxies can lose their gas reservoirs via different mechanisms, eventually quenching their star formation. We explore the diverse environmental histories of galaxies through a multiwavelength study of the combined effect of ram-pressure stripping and group `processing' in Abell 963, a massive growing cluster at $z=0.2$ from the Blind Ultra Deep \HI\ Environmental Survey (BUDHIES). We incorporate hundreds of new optical redshifts (giving a total of 566 cluster members), as well as Subaru and \textit{XMM-Newton} data from LoCuSS, to identify substructures and evaluate galaxy morphology, star formation activity, and \HI\ content (via \HI\ deficiencies and stacking) out to $3\times R_{200}$. We find that Abell 963 is being fed by at least seven groups, that contribute to the large number of passive galaxies  outside the cluster core. More massive groups have a higher fraction of passive and \HI\ -poor galaxies, while low-mass groups host younger (often interacting) galaxies. For cluster galaxies not associated with groups we corroborate our previous finding that \HI\ gas (if any) is significantly stripped via ram-pressure during their first passage through the intracluster medium, and find mild evidence for a starburst associated with this event. In addition, we find an overabundance of morphologically peculiar and/or star-forming galaxies near the cluster core. We speculate that these arise as groups pass through the cluster (post-processing). Our study highlights the importance of environmental quenching and the complexity added by evolving environments. 
\end{abstract}


\section{Introduction}
\label{sec:introduction}

It has been established by numerous observational efforts that star formation
in galaxies has steeply declined since $z\sim2$ \citep[e.g.][]{Bell2005,Daddi2007,Madau2014}. 
The declining star formation activity is accompanied by a decrease in the fraction of late-type galaxies and a comparative increase of the early-type galaxy population  \citep{Desai2007,Poggianti2009} that is particularly visible in clusters \citep[][]{Dressler1997,Fasano2000}. 
In fact, studies of large samples of nearby galaxies show a strong correlation between the passive galaxy fraction and both stellar mass and environment \citep[][]{Baldry2006,Haines2007,Peng2010}.  
Such findings have suggested that, the star formation activity of galaxies, as well as their morphologies, are strongly correlated with stellar mass, and that the quenching of star formation is accelerated in dense environments such as galaxy clusters. All of the aforementioned observations are consistent with a scenario in which galaxies are quenched as clusters grow hierarchically with cosmic time \citep[e.g.][]{deLucia2012}, and many galaxies transition from the field into groups and clusters. 
What remains unclear is which are the physical mechanisms driving galaxy evolution in different environments. 
The most likely mechanisms at play in groups and clusters \citep[see][for a review]{BoselliGavazzi06} can be divided in three broad categories.   
%
\begin{itemize}
  \item \textit{Gravitational interactions among galaxies. } 
  Simulations have shown that a major merger between two disc galaxies can result in a massive elliptical \citep[][]{ToomreToomre1972,BarnesHernquist1996}, while the accretion of small satellites by a disc galaxy can create an S0 \citep[][]{Walker1996}. 
  Galaxy-galaxy interactions however, 
  are expected to be more frequent in low-mass groups, or the field, where the relative velocities are low enough for the collisions to occur. 
  In clusters, galaxies can instead experience fast encounters with other cluster members. The accumulation of such encounters, known as \textit{harassment} \citep[][]{Moore1999}, can cause disc thickening and gas fuelling of the central region. Harassment can also strip \HI\ from galaxies \citep{DucBournaud2008}, although it is most effective in low-surface-brightness galaxies, and towards the core of galaxy clusters \citep[][]{Moore1999,Smith2015}.

  \item \textit{Galaxy-Cluster interaction.} The gravitational potential of a cluster can affect the cluster galaxies, by, for example, compressing the gas, inducing inflow, forming bars, and centrally concentrating the star formation \citep[see e.g.][]{Miller1986,Byrd1990}. However, \citet{BoselliGavazzi06} showed that gas is hardly directly removed in such interactions. 
  \item \textit{Interaction between galaxies and the intracluster medium (ICM).} \citet{GunnGott1972} first predicted (analytically) that \textit{ram-pressure} can effectively remove a galaxy's interstellar medium (ISM) as it passes through a dense ICM, thereby quenching its star formation. The efficiency of the stripping is proportional to the density of the ICM ($\rho_{\rm ICM}$) and to the square of the galaxy's velocity ($v_{\rm gal}^2$). When the ram-pressure is greater than the anchoring force of the galaxy, the diffuse ISM will be stripped \citep[][]{Abadi1999,Quilis2000,Vollmer2001,Jaffe2015}. 
  Some models predict that \HI\ can be fully depleted after a single transit through the ICM's core \citep{Roediger2007}. 
  It is also possible that only the diffuse hot halo gas can be ram-pressure stripped or thermally evaporated %
  \citep{Bekki2002}. This process, known as \textit{starvation} or \textit{strangulation}, halts the funnelling of gas into the disc, provoking a slow decline of the galaxies' star formation activity. The quenching time-scale of starvation ($\sim 4$~Gyr) is much longer than that of ram-pressure stripping (RPS) of the disc gas \citep[see][and references therein]{Peng2015}.

\end{itemize}

It is possible that at least some of the aforementioned mechanisms act at the same time, increasing the difficulty of separating them observationally. One complication is that clusters continuously grow via the accretion of galaxies from the field, filaments and/or groups. In fact, it has been proposed that galaxies can start to be quenched in low-mass groups, that ``pre-process'' them (via galaxy-galaxy interactions, galaxy-group tidal interactions, and/or RPS) before falling into a more massive cluster \citep[e.g.][]{ZabludoffMulchaey1998, Balogh1999,VerdesMontenegro2001,Lewis2002,McGee2009,deLucia2012,Jaffe2012}. 
Although the relative importance of group pre-processing in the quenching of cluster galaxies is still debated, it has been estimated from cosmological simulations that
a $10^{15} h^{-1} M\odot$ cluster at $z=0$ has accreted $\sim40\%$ of its galaxies ($M_{\bigstar}>10^9 h^{-1} M\odot$) from groups more massive than $10^{13} h^{-1} M{\odot}$ \citep{McGee2009}. Moreover, recent observations show that around half of galaxies in groups, and a significant fraction of galaxies in filaments show signs of  quenched star formation activity \citep[][]{Cybulski2014}. 

In addition, the simulations of \citet[][]{Vijayaraghavan2013}  showed  that during a group's pericentric passage within a cluster, galaxy-galaxy collisions are enhanced. Shock waves produced by the merger also enhance the ram-pressure intensity on group and cluster galaxies, while the increased local density during the merger leads to greater galactic tidal truncation. The combination of these effects has been referred to as  group `post-processing'. 
Observations suggest that $10-20$\% of clusters at $z<0.3$ are undergoing mergers with other clusters \citep{Katayama2003, Sanderson2009, Hudson2010}, and this fraction is expected to increase with redshift \citep{Mann2012}. Most studies find that cluster mergers trigger star formation, although there has been some debate \citep[][and references therein]{Stroe2015}. 
  
Neutral hydrogen (\HI\ ) is an ideal tool to study environmental processes as it is a very fragile component of galaxies \citep{HaynesGiovanelli1984,GiovanelliHaynes1985}. 
In fact, there is unimpeachable evidence that \HI\ can be disturbed and stripped in late-type galaxies as they pass through the dense ICM of massive galaxy clusters  \citep[e.g.][]{Cayatte1990,BravoAlfaro2000,BravoAlfaro2001,PoggiantiVG2001,Kenney2004,crowl05a,Chung2007,Chung2009,Abramson2011,Scott2010,Scott2012,Gavazzi2013} and even groups  \citep[e.g.][]{Catinella2013, Hess2013}. 
Simulations suggest that this happens via RPS \citep[see][for a review]{Roediger2009} and/or gravitational interactions \citep[e.g. ][]{Vollmer2001,Vollmer2003,Kapferer2009,TonnesenBryan2009}. 
Furthermore, \HI\ is also known to be sensitive to galaxy-galaxy interactions \citep[see e.g.][]{Yun1994}, common in group environment. 
Therefore, \HI\ is the perfect tool to study the effect of environment on galaxies across cosmic time. 

Due to technical limitations \HI\ has not been studied beyond the low-redshift Universe. The Blind Ultra Deep \HI\ Environmental Survey \citep[BUDHIES;][]{Verheijen2007,Verheijen2010,Jaffe2013}, observed, for the first time, \HI\ in and around two clusters at $z\sim0.2$ with the Westerbork Synthesis Radio Telescope (WSRT) detecting 160 galaxies. Recently, this redshift regime has started to be explored by other \HI\ surveys. 
The HIGHz Arecibo survey observed 39 galaxies at $0.16 < z < 0.25$ in pointed observations \citep{Catinella2015}. 
CHILES, the COSMOS \HI\ Large Extragalactic Survey is currently observing a a $40\times40$arcmin field  with the Very Large Array (VLA) covering the range  $0<z<0.5$. A pilot for this survey covering $0< z<0.193$ detected 33 galaxies  \citep{Fernandez2013}.

In \citet{Jaffe2012} we studied the \HI\ content in Abell 2192 \citep[A2192\_1 hereafter, following the notation of][]{Jaffe2013},  a cluster with mass$\sim 2\times10^{14} M\odot$, and a significant amount of substructure.  We found that this intermediate-size cluster is in the process of assembling, and that the incidence of \HI\ -detections significantly correlates with its substructure. At large clustercentric radii ($\gtrsim $2-3 virial radii), where the infalling substructures were found, many galaxies are \HI\ -rich, while at the core of the forming cluster, none of the galaxies are \HI\ -detected. The reduction of \HI\ starts to become significant in low-mass groups (with masses as low as $\sim 10^{13}M\odot$), so that  by the time the group galaxies fall into the cluster their \HI\ reservoir is significantly reduced. 

In \citet{Jaffe2015} we concentrated on the more massive cluster Abell 963 (A963\_1 hereafter, following previous papers)  to study the effect of RPS by the ICM. We  compared the location of \HI\ -detected cluster galaxies in velocity versus position 
phase-space diagrams with
the distribution of dark matter haloes in 
zoom-in cosmological simulations, and found strong evidence for \HI\ gas stripping (down to the detection limit of BUDHIES) on first infall into the cluster. In particular, we defined regions in phase-space where galaxies are expected to be experiencing \HI\ stripping via ram-pressure. We also found however that ram-pressure cannot be the only mechanism at play, since at the outskirts (infall region) of the cluster there is a significant fraction of passive \HI\ -poor galaxies. We speculated that these galaxies had to have been quenched prior to entering the cluster. 

\begin{figure}
\centering
\includegraphics[trim=0 0 0 0,scale=0.54]{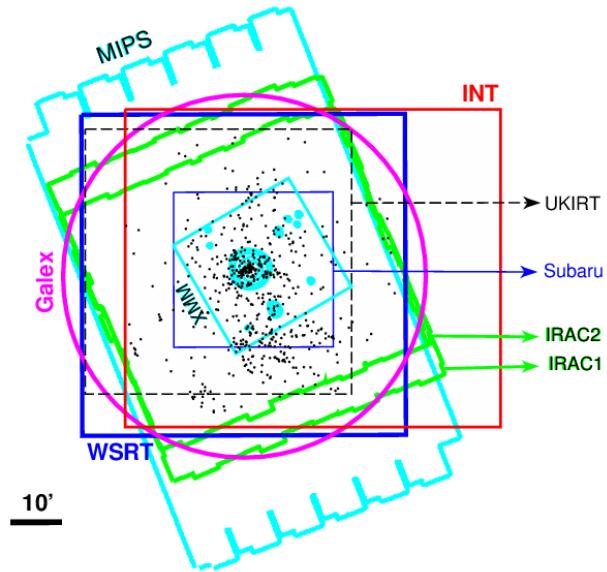}
\caption{Coverage of the different observations around the central galaxy of A963\_1. The field-of-view of the WSRT \HI\ observations is highlighted with a thick blue box and labelled accordingly, along with the multiwavelength photometry: INT ($B$ and $R$; red square), \textit{Spitzer} MIPS (cyan) and IRAC (green regions),  \textit{XMM-Newton} (X-rays; cyan tilted square), \textit{Galex}  ($NUV$ and $FUV$; magenta circle), Subaru  ($Ic$ and $V$; thin blue square), and UKIRT (dashed black square). 
In addition, the black dots show the distribution of all the spectroscopic members of A963\_1 (from WHT and MMT) and the cyan contours the extended X-ray emission detected by XMM. The horizontal black line at the bottom-left corner of the plot indicates the scale.  
 \label{fovs}}
\end{figure}

In this paper, we combine BUDHIES data around A963\_1 with hundreds of new redshifts, as well as optical and X-ray imaging data from the  Local Cluster Substructure Survey (LoCuSS) to pursue an in-depth investigation of the combined effect of RPS by the ICM with group pre- and post-processing in the removal of \HI\ and subsequent star formation enhancement/quenching in cluster galaxies.  

The paper is organized as follows: in Section~\ref{sec:data} we present the multiwavelength data available in the field of A963\_1.  
In Section~\ref{subsec:substructures} we study the degree of substructures within the cluster and identify infalling groups. In Sections~\ref{subsec:properties}~and~\ref{subsec:morphologies} we explore the properties of the cluster members, such as  galaxy colours  and morphologies as a function of environment (i.e. position within the cluster, groups, and phase-space distribution). 
In Section~\ref{subsec:HIstacks} we study the \HI\ content of galaxies as a function of environment through \HI\ deficiencies and \HI\ stacking. We complement our results in Section~\ref{subsec:Opt_stacks} by performing stacking of optical spectra in different regions of phase-space.
Finally, in Section~\ref{sec:conclu} we discuss and summarize our results. 

Throughout this paper we use AB magnitudes, and assume a concordance $\Lambda$CDM cosmology with $\Omega_{\rm M}=$0.3, $\Omega_{\Lambda}=$0.7, and H$_{0}=$70 km s$^{-1}$ Mpc$^{-1}$.


\section{DATA}
\label{sec:data}

\subsection{BUDHIES}
\label{subsec:budhies}

This study is based on data from BUDHIES, an ultra-deep \HI\ survey of galaxies in and around two clusters at $0.16\leq z\leq0.22$. 
With  effective volume depth of 328 Mpc and a coverage on the sky of $\sim$12$\times$12 Mpc for each cluster, BUDHIES covers a large range of environments. 
The ultra-deep \HI\ observations, taken with the WSRT \citep[][]{Verheijen2007,Verheijen2010}, yielded 126 \HI\ detections in the volume of A963\_1 and 36 in A2192\_1 (Verheijen et al., in preparation). The synthesized beam has a size of $23\times37$ arcsec$^2$ (corresponding to $76\times122$ kpc$^2$ at $z=0.2$), and the velocity resolution is 19 km~s$^{-1}$, allowing the \HI\ emission lines to be well resolved. 
Four \HI\ detection criteria were applied to data cubes of different resolution, namely, that the \HI\ signal was above the rms noise in a number of consecutive channels: $4\times3$; $3\times4$;  $2\times5$; or  $1\times8$ $\sigma \times {\rm d}v$, where ${\rm d}v$ is the velocity resolution element. The \HI\ mass ($M_{\HI\ }$) detection limit  was typically $M_{\HI\ }\sim2\times10^9\rm M_{\sun}$, although it varies with distance from the pointing centre (due to primary beam attenuation) and frequency.

BUDHIES also has $B$- and $R$ images from the Isaac Newton Telescope (INT), \textit{GALEX} $NUV$ and $FUV$, \textit{Spitzer} IRAC and MIPS, 
UKIRT in the near-infrared, and \textit{Herschel} SPIRE and PACS for a limited region. 
SDSS photometry is also available for the galaxies in the field, as well as spectroscopy, although this is scarce at $z\sim0.2$. 
To alleviate the lack of redshifts in the surveyed volumes, we pursued a spectroscopic campaign with WIYN and the AF2 on the 4.2m \textit{William Herschel Telescope} (WHT), from which we have assessed cluster memberships, measured the equivalent width of the [O$_{\rm II}$] line when present, and characterized the environment in the surveyed volumes \citep[][]{Jaffe2013}. The spectroscopic campaign had a spectral coverage of 3900--6900{\AA} and limited completeness ($>40$\% down to $R<19.5$ in A963\_1, the least complete cluster in BUDHIES).  
To increase the spectroscopic completeness in and around A963\_1, in this paper we have included hundreds of new redshifts from LoCuSS (private communication) and \citet{Hwang2014}. 
As part of the LoCuSS cluster sample A963\_1 was targeted with Hectospec, a multi-object spectrograph on the 6.5m MMT telescope. 
The observations provide a wide wavelength range (3650--9200{\AA}) at 6.2{\AA} resolution. Targets were selected from UKIRT/WFCAM $K$-band images down to $K{=}16.964$, with $J{-}K/K$ colours consistent with known cluster members over a $52 {\rm arcmin}{\times} 52{\rm arcmin}$ field of view \citep{Haines2009b}, and additional preference was given to those targets detected at 24$\mu$m \citep{Haines2013}. 
\citet{Hwang2014} also obtained more redshifts  in the volume of A963\_1 using 
MMT/Hectospec.  They targeted galaxies brighter than $m_{r,{\rm Petro},0}=20.5$ in the field of the cluster, irrespective of colour. 
The inclusion of all new redshifts have increased our spectroscopic completeness significantly (to above $70\%$ at $R<19.5$), as shown in Fig.~\ref{completeness} of Appendix~\ref{sec:newz}.

In addition to the spectroscopic observations, LoCuSS also observed the inner ($<R_{200}$) region of A963\_1 with \textit{XMM-Newton} \citep{Martino2014} and Subaru/Suprime-Cam imaging  in $V$ and $Ic$ \citep[with a seeing of $0.75$~arcsec, corresponding to 2.5 kpc at the cluster redshift; see][]{Okabe2010}.  We make use of these data sets in Section~\ref{subsec:A963} to study cluster substructure, and in Section~\ref{subsec:morphologies} to characterize the morphologies of the cluster galaxies.

Fig.~\ref{fovs} shows the coverage of the multiwavelength observations in the field of A963\_1 used in this paper, along with the spatial distribution of the cluster members. 

\begin{figure}
\centering
\includegraphics[trim=70 20 40 15, scale=0.54]{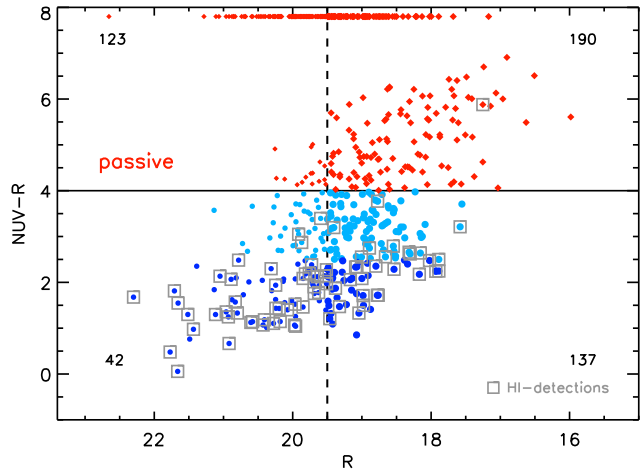}
\caption{ CMD of all the cluster members of A963\_1. 
The horizontal line divides galaxies into ``passive'' ($NUV-R>4.0$ or undetected in $NUV$, plotted at $NUV-R=7.8$ at the top of the plot for reference; red diamonds), and blue ($2.5<NUV-R<4.0$ and $NUV-R<2.5$; light and dark blue circles respectively). 
The vertical dashed line shows the spectroscopic completeness limit $R=19.5$). 
We use different symbol shapes, colours and sizes in this plot to easily identify each sample (e.g. passive versus star-forming or  bright versus faint) in Fig.~\ref{radec_col}. 
The number of galaxies in each CMD region are labelled, and \HI\ -detections are marked with open grey squares.  
 \label{CMD}}
\end{figure}

\begin{figure*}
\centering
\includegraphics[trim=20 10 20 10,scale=0.53]{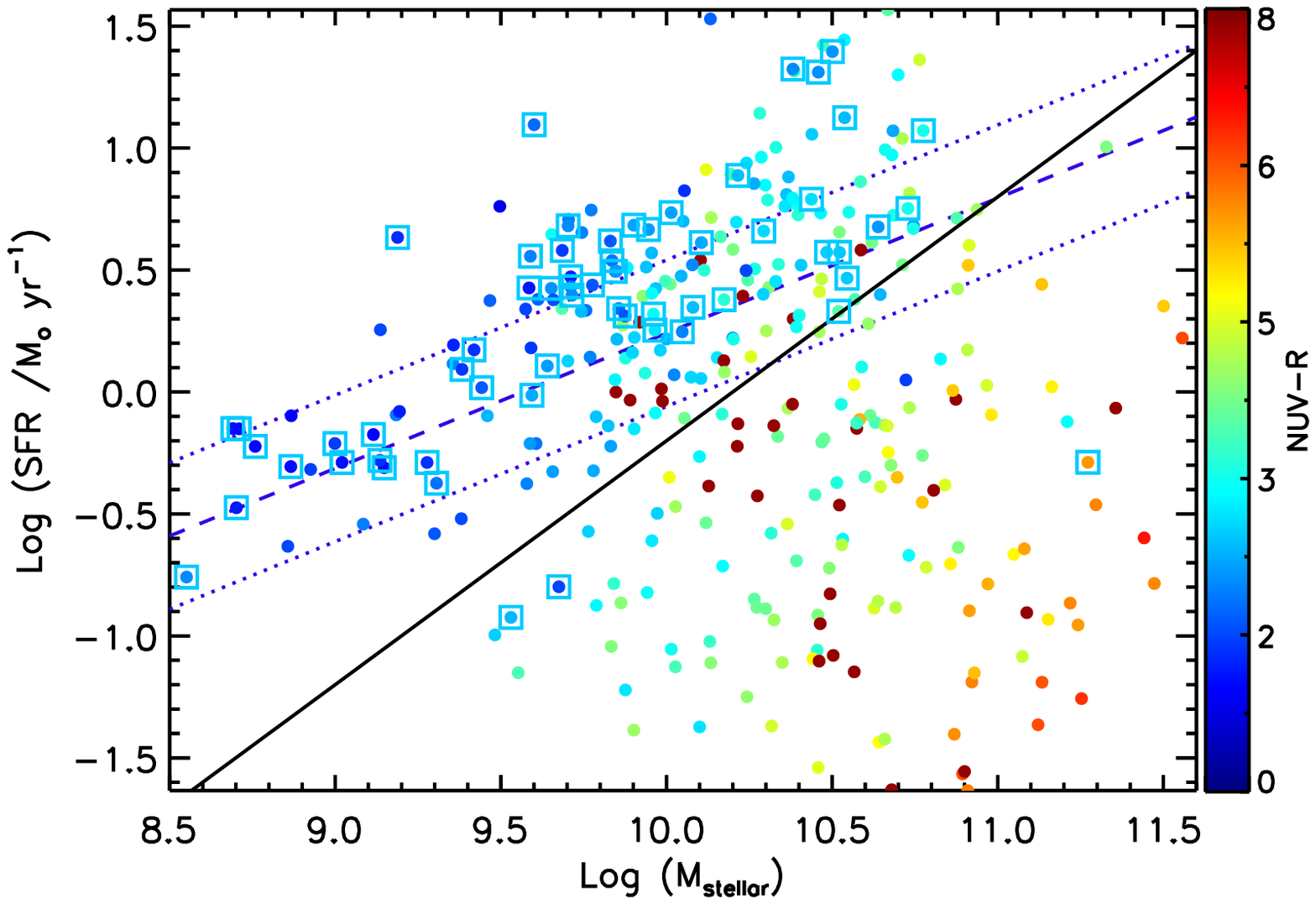}
\includegraphics[trim=14 10 20 10,scale=0.53]{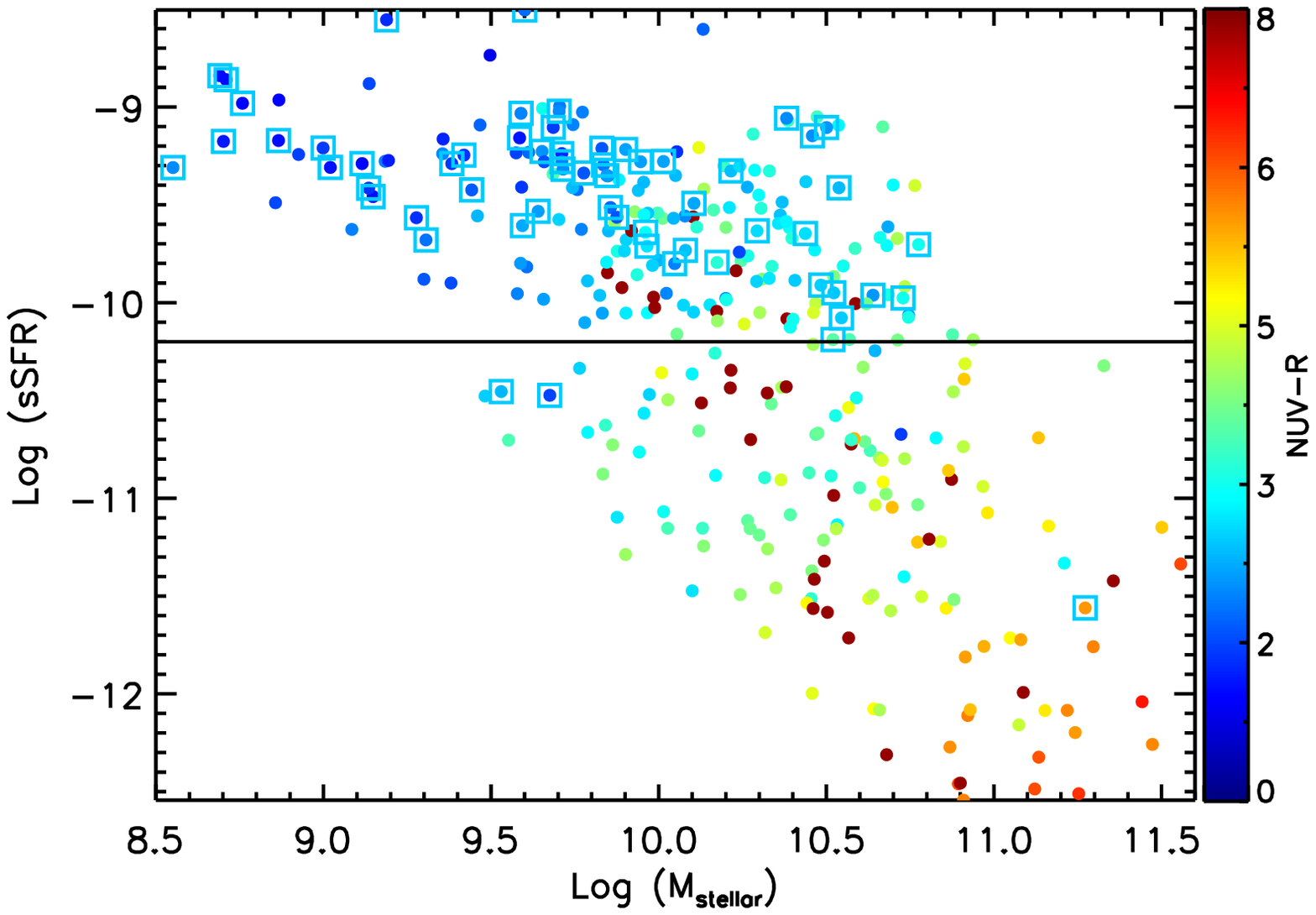}
\caption{SFR (left) and sSFR(right) versus stellar mass for all cluster galaxies, colour-coded by $NUV-R$ colour. Galaxies not detected in NUV were plotted with the reddest colour of the scale. In both panels, \HI\ -detected galaxies are highlighted with open cyan squares. 
The star-forming ``sequence'' at $z=0.2$ from \citet{Speagle2014} is plotted for reference with a blue dashed line (and dotted lines for scatter). The solid line (in both panels) corresponds to sSFR$=10^{-10.2} {\rm yr}^{-1}$, and 
roughly separates blue \HI\ -detected galaxies from redder undetected ones. 
 \label{sfs}}
\end{figure*}

\subsection{A963\_1}
\label{subsec:A963}

This paper focuses on A963\_1 (RA$=$10h 27m 36.97s; DEC$=$-9d 18m 11.34s), the most massive cluster of BUDHIES. 
The improved completeness, and the additional 426 galaxies with a redshift allowed us to better characterize the mass of the cluster and the substructure within it (Section~\ref{subsec:substructures}). From these new estimates we also re-assigned cluster memberships considering both their position and velocity, as can be seen in  the right-hand side of Fig.~\ref{LSS}. The figure shows a projected clustercentric distance versus redshift phase-space diagram in an extended field around the BUDHIES volume. The cluster members are coloured in red, and were chosen to follow the ``trumpet'' shape characterizing the cluster. 
The revised mean redshift for the cluster is $z_{cl}=0.2042$.

From a sample of 140 spectroscopic cluster members ($R<19.5$), in \citet{Jaffe2013} we estimated the cluster's mass to be $M_{200}=1.6 \times 10^{15} M\odot$ ($\sigma_{cl}= 993$ km s$^{-1}$). With the increased spectroscopic sample we now have 566 cluster members. Using the same magnitude-limit as in  \citet[][]{Jaffe2013}  we  re-computed the mass of the cluster to be: 
$1.41 \times 10^{15} M\odot$ ($\sigma_{cl}= 970$ km s$^{-1}$). This is still somewhat higher than the most recent X-ray (\textit{XMM-Newton}) $M_{200}$ estimate, $9.17 \pm 2.07 \times 10^{14} M\odot$\footnote{ 
This value was measured by  extending the profiles in \citep{Martino2014} to $R_{200}=1.853 +/- 0.145 Mpc$  (Haines et al. in preparation)}, but closer to the latest weak lensing mass estimate \citep[$10.39^{+2.11}_{-1.19} \times 10^{14}M\odot$][] {OkabeSmith2015}. 
We note however that our mass estimate (inferred from $\sigma_{cl}$) assumes that the cluster is relaxed and in equilibrium. The discrepancy between the measurements could thus be caused by the presence of substructure. 

A963\_1 has been classified as regular by previous X-ray studies and lensing analysis \citep{Smith2005,Chon2012}. However, in \citet{Jaffe2013} we saw weak evidence for substructure, based on a kinematic analysis. In Section~\ref{subsec:substructures} we will indeed show that there are several distinct groups feeding the cluster. 
If we exclude galaxies belonging to subgroups, the dynamical mass drops down to $1.15 \times 10^{15}M\odot$ ($\sigma=907$ km s $^{-1}$), which coincides with the weak lensing mass estimate. 

\subsection{Colours, stellar masses and star formation rates}
\label{subsec:sfr_and_mass}

Fig.~\ref{CMD} shows the 
$NUV-R$ versus $R$ colour-magnitude diagram (CMD) of the spectroscopically confirmed members of A963\_1. 
Because $NUV-R$  traces the ratio of young to old stars, red $NUV-R$ colour can  be associated with passive galaxies. 
Galaxies not detected in $NUV$ are considered ``red'' and are plotted in Fig.~\ref{CMD} with arbitrarily red colour ($NUV-R$=7.8, redder than the reddest cluster galaxy) for reference. 
The vast majority of the undetected galaxies lie in the $B-R$ versus $R$ red-sequence and do not show emission in their optical spectra. We therefore refer to all galaxies in the red ($NUV-R>4.0$) bin of the CMD, including galaxies not-detected in $NUV$ as ``passive'' hereafter.  
In the figure, \HI\ -detections  are also highlighted (squares) and the number of galaxies in the different samples is indicated. Most of the \HI\ detections are in the bluest part of the diagram, as expected. In particular, it is visible how the \HI\ detections drop with $NUV-R$ colour so that almost no galaxy redder than $NUV-R\sim3$ is detected in \HI\. Our findings are consistent with the work  of~\citet{Cortese2011}, \citet{Catinella2013} and \citet{Brown2015}, that show a gradual decrease in \HI\ content of $z=0$ galaxies with $NUV-R$ colour. The only exception is one passive galaxy detected in HI, which is an elliptical galaxy with a small nearby galaxy (in projection) that could be responsible for the \HI\ emission. 
The CMD of Fig.~\ref{CMD} shows a prominent population of dwarf blue cluster galaxies that extend 1-2 mag fainter than red dwarfs. This reflects the BUDHIES spectroscopic target selection criteria that allocated fibers to galaxies brighter than $R=19.5$ with the exception of \HI\ -detected galaxies that could be as faint as $R=21$ \citep[see Fig. 2 in][]{Jaffe2013}. Also, for the faint population, spectroscopic redshifts were easier to obtain when emission-lines were present. For $R<19.5$ however, our sample is fairly unbiased and complete (see Figure \ref{completeness}).

It is known that the $24\mu$m emission reflects the obscured star formation \citep{Kennicutt2007}, while the observed UV emission traces the un-obscured component  \citep{Calzetti2007}. 
We thus computed total star formation rates (SFR) from a combination of \emph{GALEX} FUV luminosity and infrared luminosity ($L_{\rm IR}$) derived from MIPS. First, we derive $L_{\rm IR}$ from the MIPS $24\mu$ flux and redshift using the calibration of \citet{Rieke2009}. For the un-obscured component, we calculate $L_{\rm FUV}$ after applying a $k$-correction to the FUV flux. Then, for galaxies detected in both FUV and MIPS $24\mu$, we calculate total SFR following Equation 9 of \citet{Murphy2011}. For galaxies detected only in FUV, we calculate SFR using Equation 3 of \citet{Murphy2011}, which accounts for only the un-obscured component. And for any galaxies detected in MIPS $24\mu$ but not in FUV, we use Equation 4 of Murphy et al. (2011). We note that the computed SFRs are dominated by the infrared component, and thus they trace the obscured component of star formation activity well. Specific star formation rates (sSFRs) were computed dividing SFR by stellar mass ($M_*$). We measured $M_{*}$ from the available $B$ and $R$ INT photometry, using the method described in \citet{Zibetti2009}, and a Kroupa initial mass function.

Fig.~\ref{sfs} shows the SFR (left) and sSFR (right) as a function of stellar mass for the cluster members, colour-coded by their $NUV-R$ colours. 
The black solid line indicates the sSFR boundary between most \HI\ -detections (that also roughly corresponds to galaxies with $NUV-R<2.5$) and non-detections. Such boundary corresponds to $sSFR\sim10^{-10.2} {\rm yr}^{-1}$, as shown in the right-hand panel. 
As evidenced by Fig.~\ref{sfs}, $NUV-R$ is sensitive to recent star formation and it can thus be used as a proxy for specific star formation rate \citep[sSFR, see][]{Salim2005, Salim2007, Schiminovich2007}.

\section{RESULTS}
\label{sec:results}

\subsection{Substructure within the cluster}
\label{subsec:substructures}

\subsubsection{X-rays}

Previous \textit{Chandra} and \textit{HST} observations of the central few hundred kpc region of the cluster reveal a regular and concentrated  X-ray emission, around a cD galaxy \citep[Fig. 6 of ][]{Smith2005}.  Recent analysis of \textit{XMM-Newton} observations by 
Haines et~al. (in preparation) has revealed additional substructures surrounding the centre, as shown by the cyan contours of Fig.~\ref{fovs} \citep[see also Fig. B.6. in][]{Zhang2007}. Three of them are at the redshift of the cluster, at a distance of $\sim R_{200}$ from the cluster centre. These cluster substructures, labelled `A', `B' and `C' and listed in Table~\ref{gr_table}, have velocity distributions consistent with that of groups, as shown in the top panel of Fig.~\ref{group_vels}.  
The most massive group detected in X-rays is group C, that could even be considered a small cluster from its mass. In fact, it has a  distinctively bright galaxy at its centre. 
This group is however over an order of magnitude less massive than A963\_1. 
Groups B and A do not display a distinctive bright central galaxy. Group B in fact hosts many merging galaxies, and the low-mass group A shows a prominent population of blue, unperturbed, late-type galaxies. 

To assign group membership we selected galaxies within a radius equal to the $R_{200}$ of the groups and line-of-sight velocities consistent with the bulk of the other group galaxies, as shown in Fig.~\ref{group_vels}.  $R_{200}$ was computed for each group from $M_{200}$, estimated using the X-ray luminosity-$M_{200}$ relation of  \citet[][see Haines et al. in preparation]{Leauthaud2010}. The $R_{200}$ of the X-ray groups are plotted as open circles in Fig.~\ref{DStest}, and labelled `A', `B' and `C'.

\subsubsection{The Dressler-Shectman test}
An alternative way to identify substructures within a galaxy cluster is through the Dressler-Shectman  \citep[DS;][]{DresslerShectman1988} test, that compares the \textit{local} (10 nearest neighbour) velocity and velocity dispersion for each galaxy with the \textit{global} (cluster) values. 

In short, the method consists of measuring individual galaxy deviations, $\delta_{i}$ \citep[Eq. 6 in][]{Jaffe2013} for the $N_{\rm mem}$ cluster members, to then quantify the degree of substructure in two ways:  
\begin{enumerate}
 \item The `critical value' method evaluates $\Delta=\sum_{i} \delta_{i}$. 
 \item The probabilities ($P$) method instead computes $P$-values by comparing the $\Delta$-value to `shuffled' $\Delta$-values $P$ is then calculated as $\sum(\Delta_{\rm shuffle} > \Delta_{\rm obs} / N_{\rm shuffle})$, so that a small value indicates the presence of substructure.
\end{enumerate}

The details of the DS test can be found in \citet{Jaffe2013}, where we carried out this analysis in all the BUDHIES clusters with the limited spectroscopic sample available at the time, finding marginal signs of substructure in A963\_1. In particular, we found weak evidence for the presence of 4 groups within the cluster (A963\_1a, A963\_1b A963\_1c and A963\_1d). 
It is known however that the mean $\Delta$-deviation substructure index is very sensitive to completeness \citep[see Fig. 3 in][]{Ragone-Figueroa2007}. 
For this reason, we re-assess here the presence of substructures using the increased spectroscopic sample.

\begin{figure*}
\centering
\includegraphics[trim=5 30 25 20,scale=0.85]{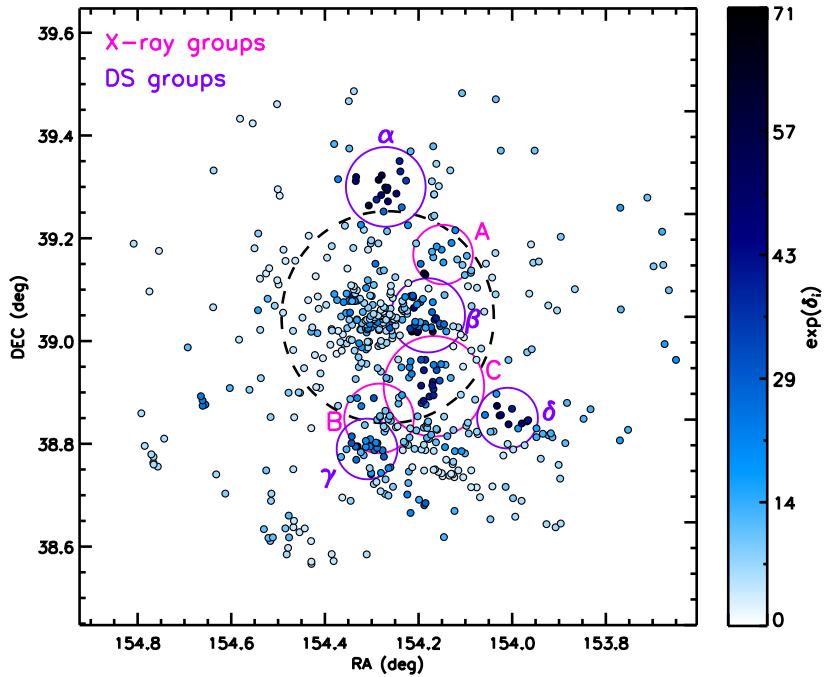}
\caption{The spatial distribution of all the spectroscopically confirmed members of A963\_1 is plotted in solid circles, colour-coded by their deviations $\delta_{i}$ (see text for details). The darker the symbols the more deviated the galaxies are from the mean velocity distribution of the cluster.  Open circles enclose the galaxy groups identified form the X-rays  (Groups A, B, and C; pink), and the DS test (Groups $\alpha$, $\beta$, $\gamma$, and $\delta$; violet), and the dashed circle corresponds to the cluster's $R_{200}$. 
\label{DStest}}
\end{figure*}

The results of the DS test on the new data reveals a significant degree of substructures ($\Delta$/$N_{\rm mem}=2.096$ and $P<0.001$). In fact,  several substructures surrounding the cluster core can be seen in Fig.~\ref{DStest}, that highlights with darker colours the galaxies with the largest deviations (i.e. the highest $\delta_i$ values). Interestingly, the X-ray groups (enclosed  by  open circles, and labelled `A', `B', and `C') do not show the strongest deviations from the local and global velocity distribution of the cluster. The strongest deviations ($e^{\delta_i} > 20$) instead define several additional substructures or galaxy overdensities that are not identified from the X-rays. The four most prominent ``DS groups'' are labelled with the Greek letters $\alpha$, $\beta$, $\gamma$ and $\delta$ for distinction from the X-ray groups, and listed in Table~\ref{gr_table}.

\begin{table}
\begin{tabular}{ccccc}
\hline
Group & $\sigma$ (km s$^{-1}$) & $M_{200}$ ($M\odot$) & No. members & Method \\
\hline
 A & 383	& $2.3\times 10^{13}$ &	10 (7) & X\\
 B & 234	& $3.2\times 10^{13}$ &	8 (6) & X\\
 C & 753	& $1.0\times 10^{14}$ &	34 (16) & X\\
\hline
 $\alpha$	& 338	&	$6.0\times 10^{13}$ 	&17 (9)&DS	\\
 $\beta$	& 1129	&	--	&22 (12)& DS	\\
 $\gamma$	& 108	&	--	&6 (5)	&DS\\
 $\delta$	& 197	&	$1.2\times 10^{13}$	&9 (4)&DS	\\
\hline
\end{tabular}
\caption{Properties of the subgroups found in A963\_1. The columns are: Group given name, velocity dispersion ($\sigma$), mass ($M_{200}$), number of members (number above the spectroscopic completeness limit), and method of identification: (X for X-rays or DS for dynamical). The observed velocity dispersion ($\sigma$) was estimated for galaxies above the spectroscopic completeness limit. 
$M_{200}$ was estimated from the X-ray data for the X-ray groups A, B, and C (top). For the DS groups (bottom) it was derived from the measured $\sigma$. We note that groups $\beta$ and $\gamma$ have very wide and narrow velocity distributions respectively, so their dynamical mass estimates were not attempted.}
\label{gr_table}
\end{table}

The DS groups $\beta$ and $\gamma$ correspond to the groups A963\_1d and A963\_1c, identified in \citet[][despite the spectroscopic incompleteness]{Jaffe2013}. Our revised analysis confirms their reality and allows us to assign secure group memberships to galaxies in these two groups. 
We also found that the group  A963\_1b in \citet[][see Fig. 10]{Jaffe2013} instead, corresponds to the X-ray group A, while group A963\_1a is not picked up as a strong DS substructure in the new DS analysis, so we discard it as a real substructure. 

\begin{figure}
\centering
\includegraphics[trim=0 -20 0 0,scale=0.4]{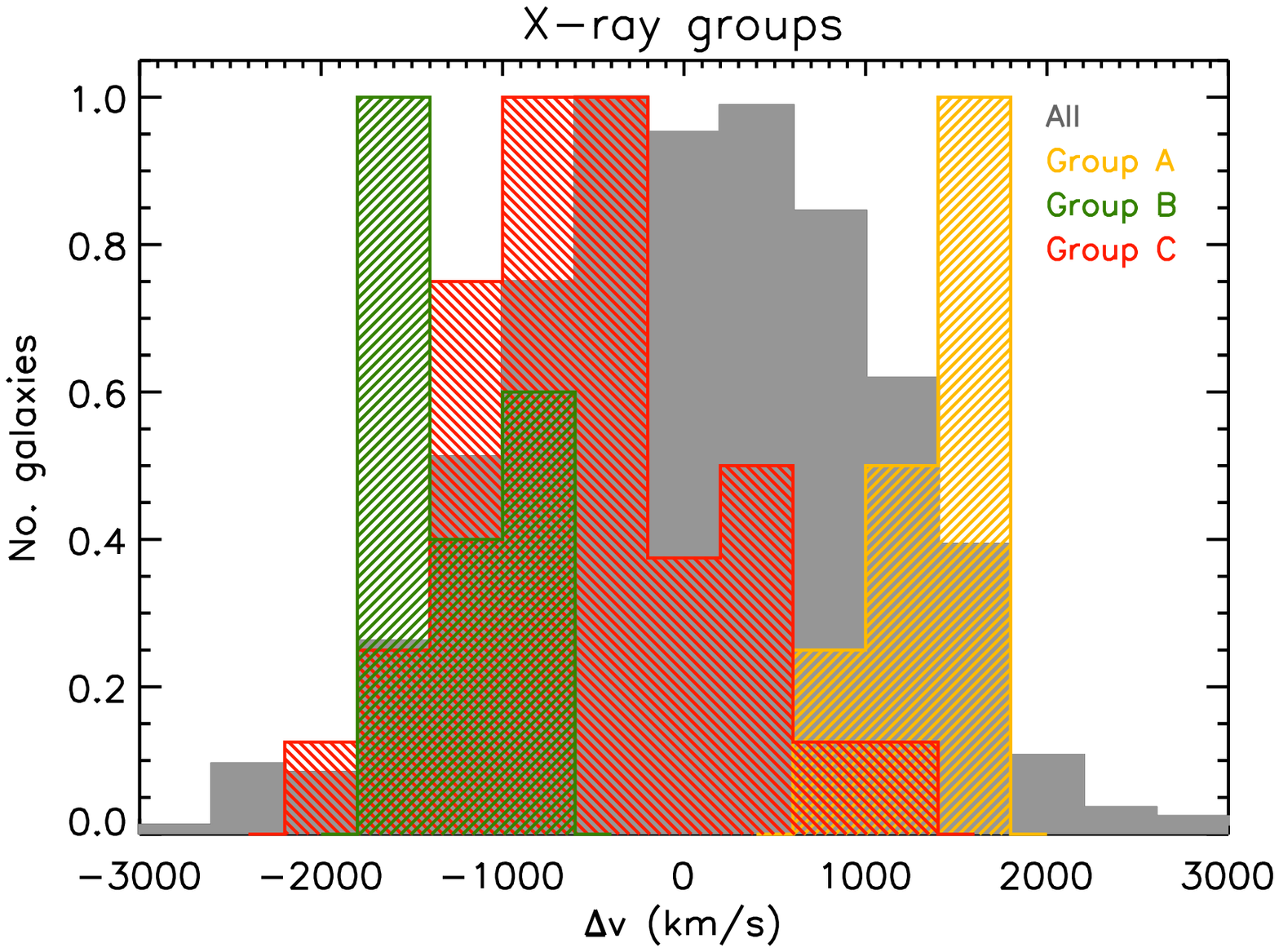}
\includegraphics[scale=0.4]{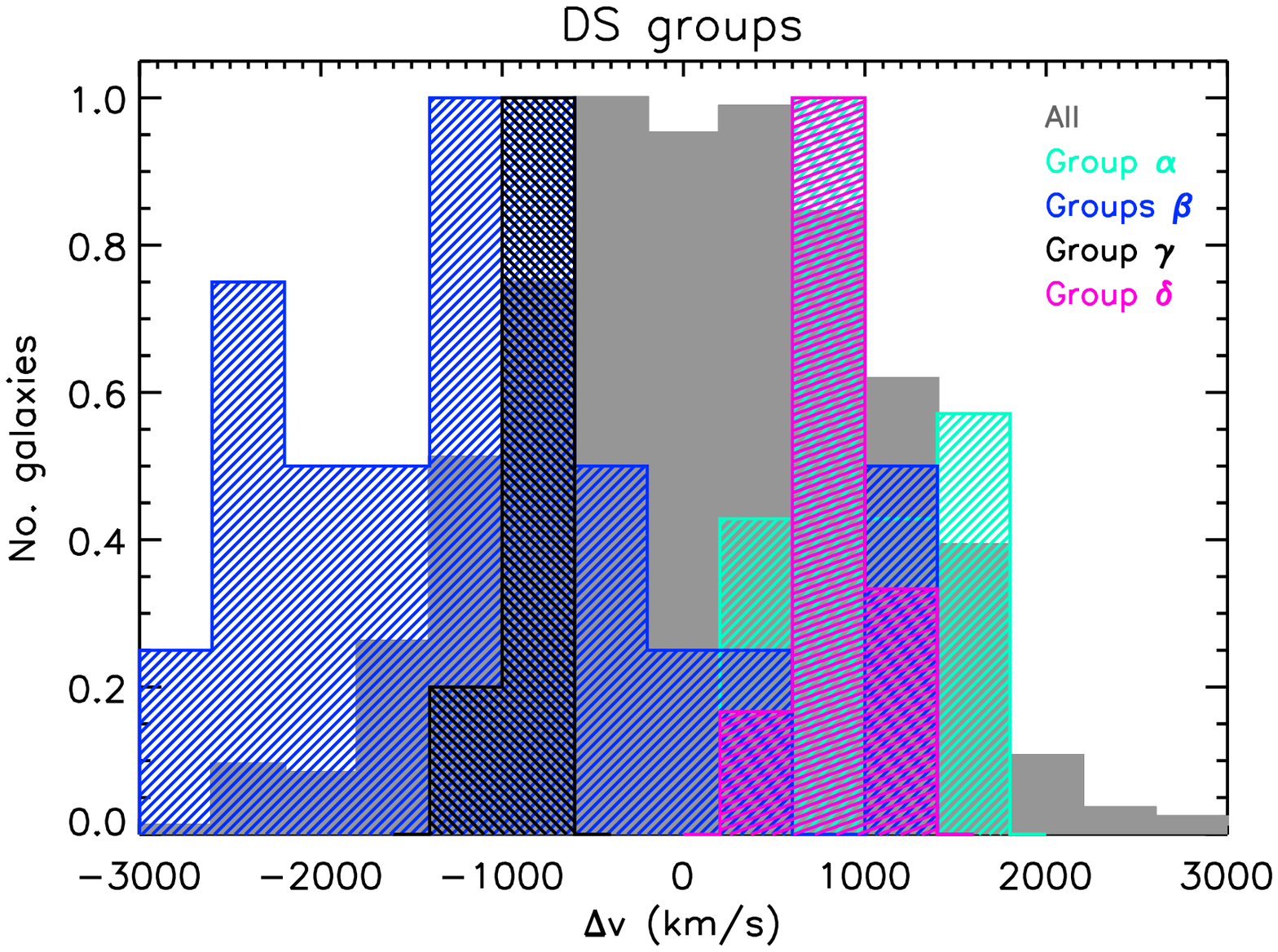}
\caption{Velocity distribution of the galaxies in the main cluster (solid grey histogram), and those in the X-ray groups A, B, and C  (left) and the DS groups $\alpha$, $\beta$, $\gamma$, and $\delta$ (right), as labelled. 
 \label{group_vels}}
\end{figure}

With the exception of group $\beta$, all the DS groups have a narrow velocity distribution as shown in the bottom panel of Fig.~\ref{group_vels}, confirming they are real associations. 
None of the DS groups are detected in X-rays, but a plausible explanation exists. 
The DS groups $\alpha$, $\delta$ and $\gamma$ are located outside of the region covered by the \textit{XMM-Newton} observations 
(see Fig.~\ref{fovs}). 
Only the group $\beta$ is well inside the \textit{XMM-Newton} region. 
It is possible however, that this group is being disrupted by the main cluster due to its vicinity to the cluster core and it is thus not detected in X-rays. This is consistent with its broad velocity distribution.  

It has been shown by \citet{Cohn2012} that it is easier to detect a substructure via the DS test for large subgroups, and that the likelihood of detection also depends on the line-of-sight direction. This could mean that the four substructures we found via the DS test in A963\_1 are the larger subgroups in A963\_1, but there could be more undetected ones. 
In fact, from Fig.~\ref{DStest} we can see signs of weaker DS structures that could be individual groups or part of disrupted groups, as will be discussed below. 

\begin{figure}
\centering
 \includegraphics[trim=20 0 10 0, scale=0.5]{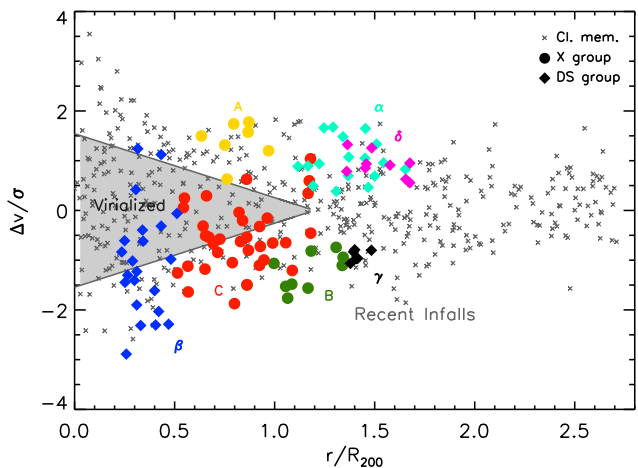}
\caption{ Projected phase-space diagram for all the cluster members  (small grey crosses). Galaxies in the X-ray or DS groups are highlighted with bigger  circles and diamonds respectively, and coloured as in Fig.~\ref{group_vels}. The `virialized' and `recent infall' zones defined in \citet[]{Jaffe2015} are indicated (see also description in Section~\ref{subsec:substructures}). 
 \label{pps_groups}}
\end{figure}

\subsubsection{Accretion history of the cluster}

\begin{figure*}
\centering
\includegraphics[trim=0 10 0 0, scale=0.511]{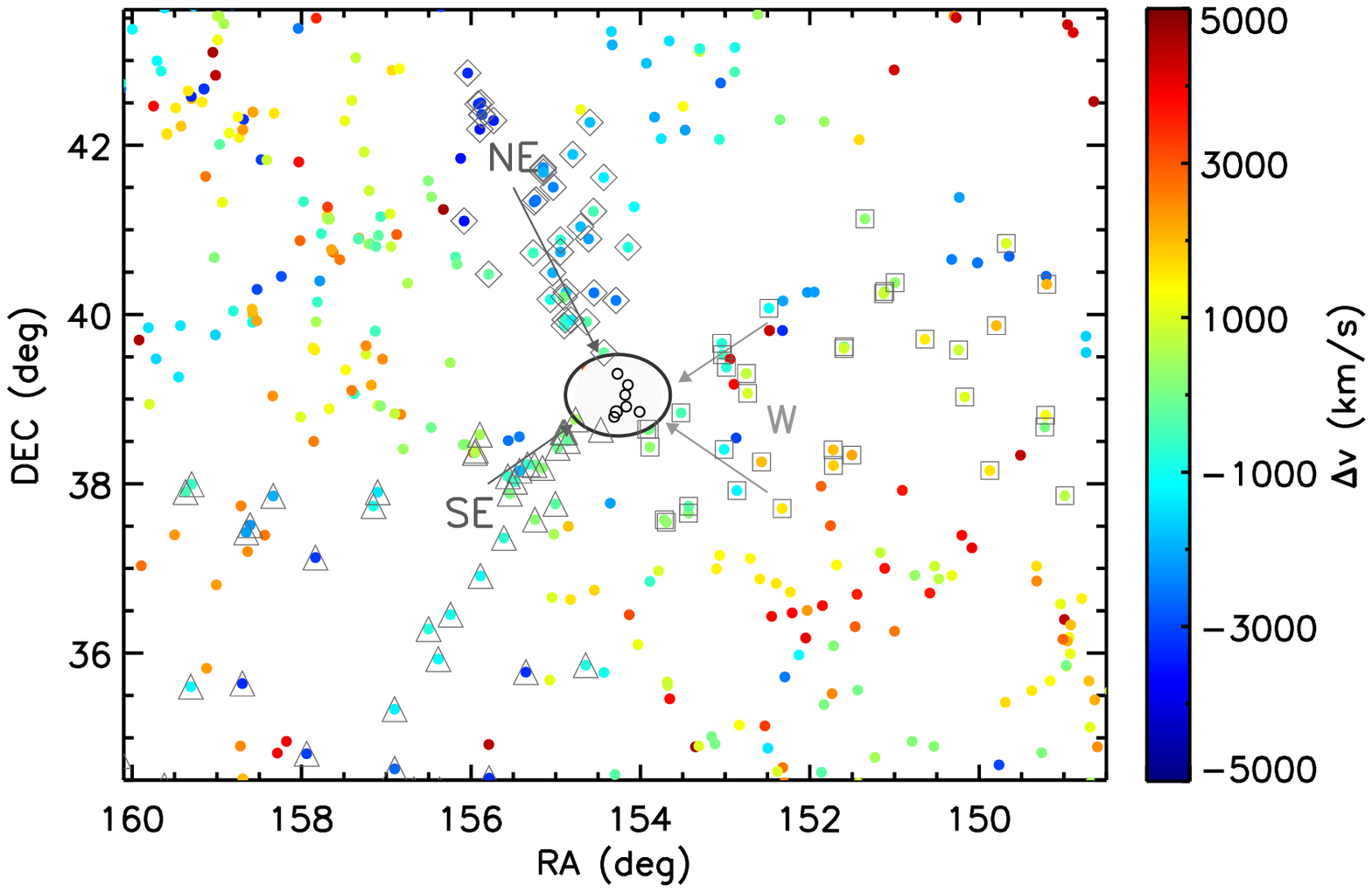}
\includegraphics[trim=0 0 0 00, scale=0.47]{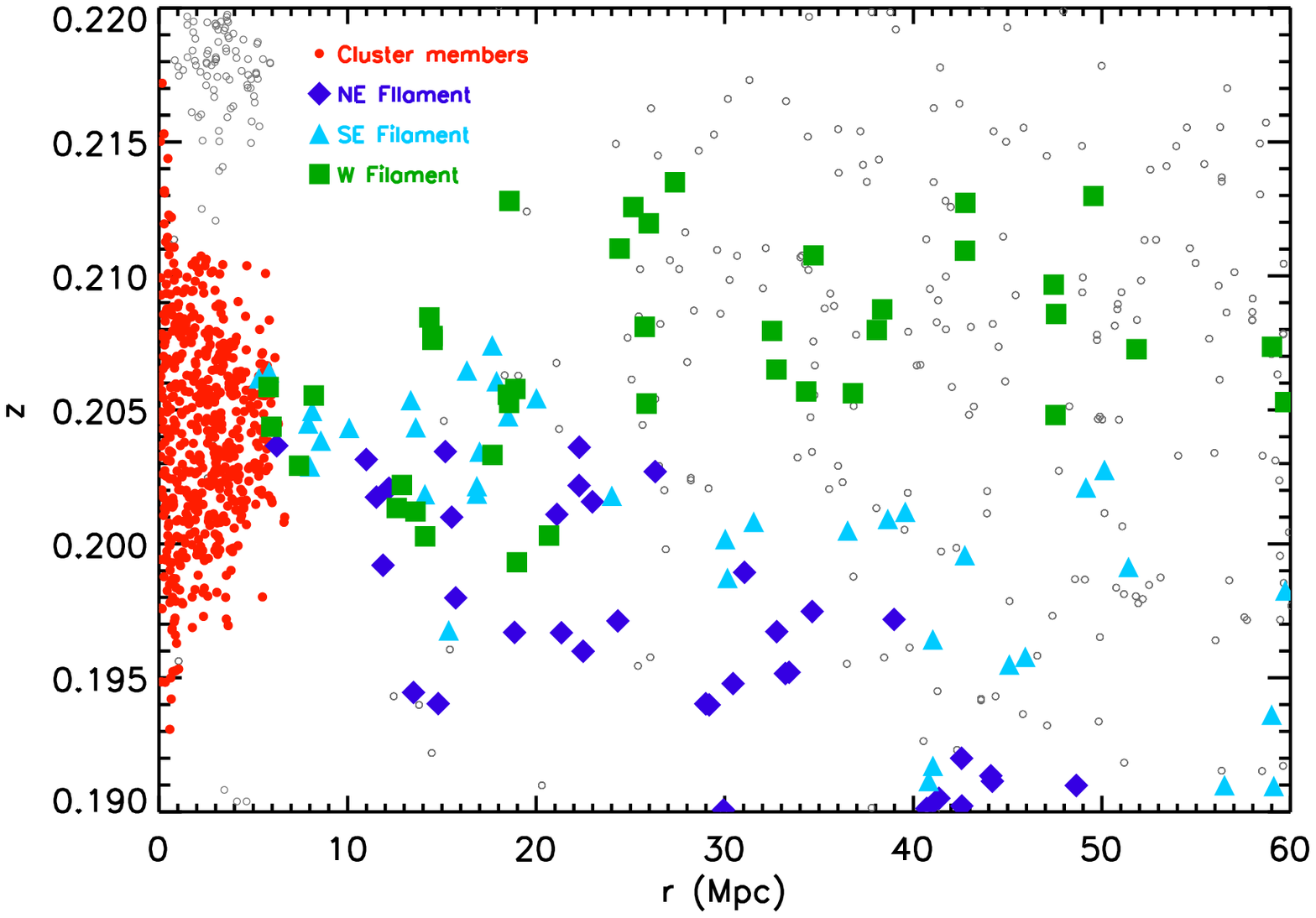}
\caption{
Left: Galaxies in the redshift range 0.184-0.224 in a $\sim9\times9$~deg$^2$ area around A963\_1. Galaxies are colour-coded according to their velocities (centred on the cluster's mean value), as indicated. The large open circle has a radius of $3\times R_{200}$ ($\backsimeq5.8 {\rm Mpc}$ at $z=0.2039$), and roughly corresponds to the coverage of our spectroscopic campaign. Outside of this circle galaxies with a redshift from SDSS are plotted. Inside of the circle, smaller open circles demark the location of the identified X-ray and DS galaxy groups. The figure shows that A963\_1 is being fed by at least two filaments (NE and SE), and possibly a third one (W). Galaxies in the filaments are enclosed by open diamonds, triangles and squares. 
Right: distance from the centre of A963\_1 (in Mpc) versus redshift for all the galaxies with a redshift between 0.190 and 0.220 (small open circles) The red filled circles are the galaxies that are considered as cluster members. Their distribution in this diagram fill the typical ``trumpet'' shape of fairly virialized clusters in phase-space (see also bottom panel of Fig.~\ref{radec_col}). The bigger symbols correspond to the galaxies in the filaments identified in the left-hand panel, as labelled.
\label{LSS}
}
\end{figure*}

The presence of sub-groups within A963\_1 is consistent with cosmological simulations \citep[e.g.][]{Cohn2012}, which predict that sub-groups host a significant fraction of cluster galaxies, with  larger  sub-groups  more  likely  to  be  in  more  massive  clusters. 
The number of galaxies in groups however, drops significantly with time since infall into the cluster.  This prediction suggests that the groups that we are able to detect inside of A963\_1 have not been in the cluster for long. Fig.~\ref{pps_groups} supports this assumption. 
The horizontal axis of the plot shows the projected distance from the cluster centre, $r$, normalized by $R_{200}$ (as computed from X-rays), while the vertical axis displays the peculiar line-of-sight velocity of each galaxy with respect to the cluster recessional velocity ($\Delta v$), normalized by the velocity dispersion of the cluster ($\sigma=907$ km s$^{-1}$; computed excluding group members). The figure also shows the `recent infall' and `virialized' regions, defined in \citet{Jaffe2015}. 
In short, the `recent infall' region represents the cluster outskirts, where galaxies that were most recently accreted into the cluster are more likely to be. 
In the `virialized' region instead, $\gtrsim70$\% of the galaxies have been in the cluster for over $\sim4$ Gyrs (roughly a cluster crossing time). 
Fig.~\ref{pps_groups} thus shows how most group galaxies clearly avoid the `virialized' region of the cluster in phase-space, preferring the `recent infall' region.

Although we assume a spherical morphology for the groups (circles in Fig.~\ref{DStest}), many of them show a non-spherical spatial distribution, such as elongations or preferred directions that can extend beyond the drawn boundaries. It is possible that some galaxies in the vicinity of these groups were previously bound to them, but are now being ejected, as the group falls deeper into the cluster's potential well. For example, there are a few galaxies with large $\delta_i$ values south of group C, west of group $\delta$ and south-west of group $\gamma$ (see Fig.~\ref{DStest}) that could have well been part of these groups in the past. 
Simulations indeed show that galaxy groups can be stretched out along the direction of infall before falling into the cluster \citep[][]{Vijayaraghavan2013}.  This could be a consequence of the group falling in along a cosmological filament or the effect 
of the cluster’s tidal field, or a combination of both.  
The elongations of some of the groups (together with their phase-space location) thus suggest that (at least some of) the groups  are on first infall and could be part of a larger filamentary structure.

\begin{figure*}
\begin{center}
\includegraphics[trim=0 0 40 0, scale=0.72]{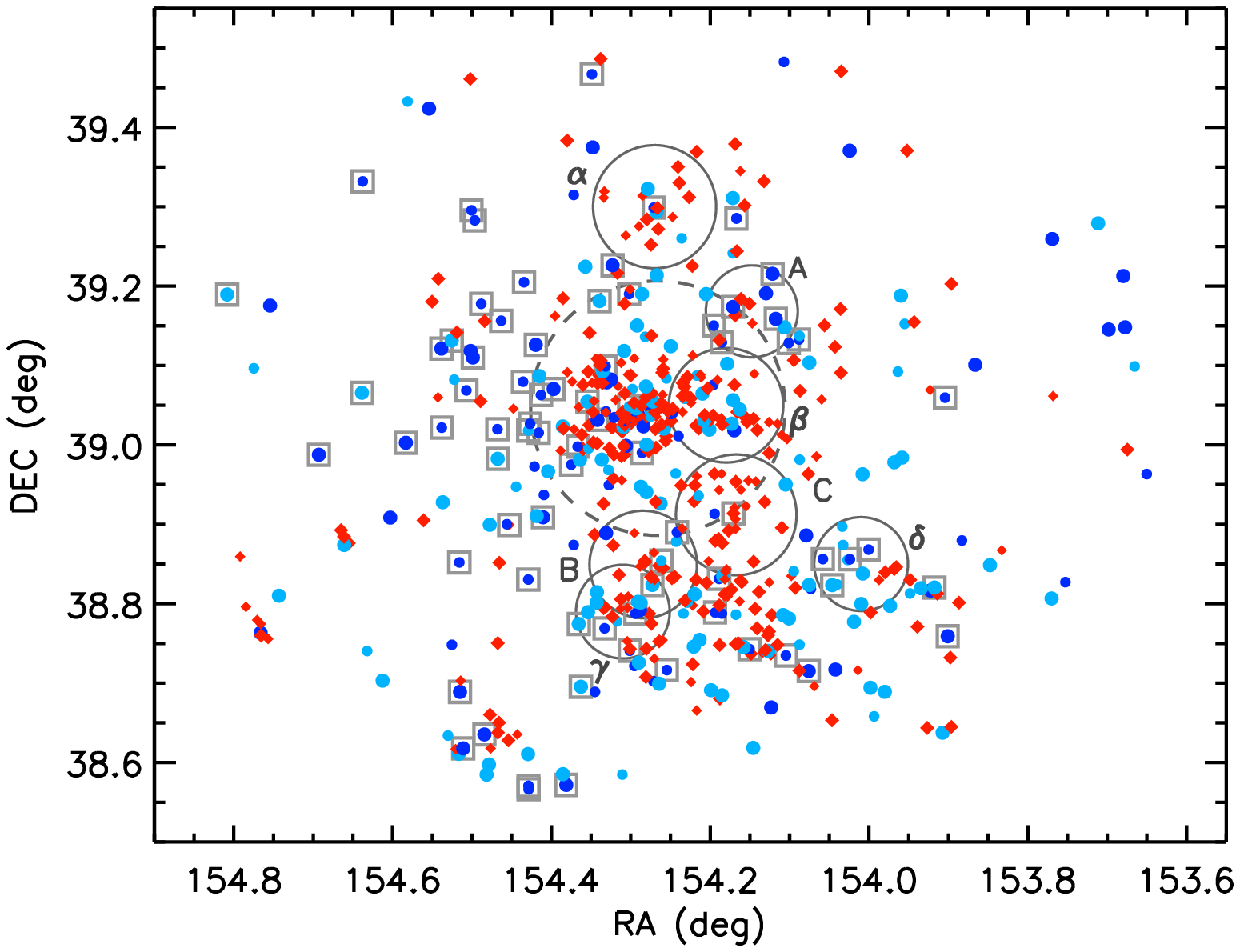}
 \includegraphics[scale=0.72]{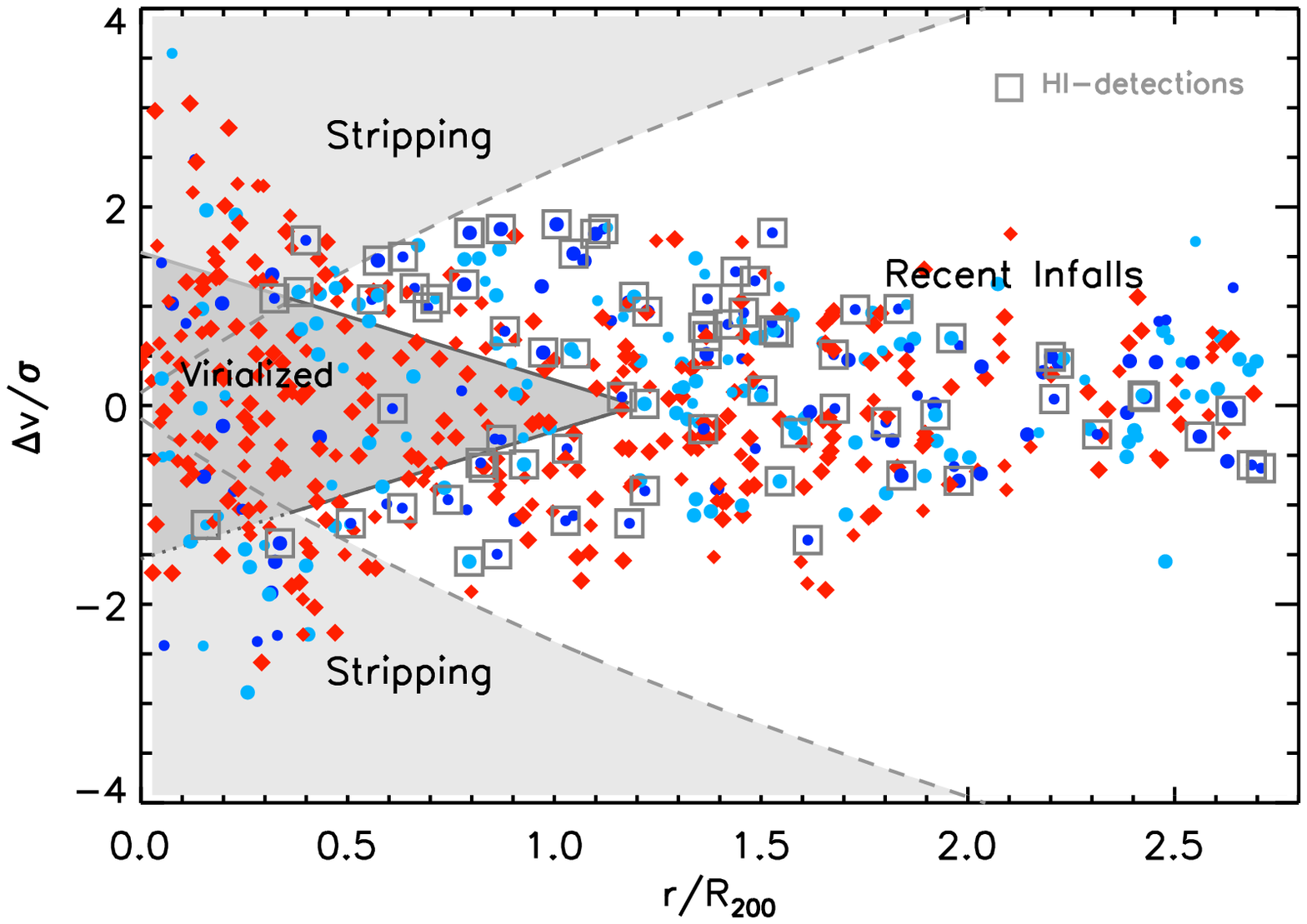}
\caption{Spatial (top) and phase-space (bottom) distribution of all the cluster galaxies, colour-coded by $NUV-R$ and sized by luminosity as in Fig.~\ref{CMD}.  The open grey squares correspond to the \HI\ -detected galaxies. Open circles in the top panel mark the groups defined in Fig.~\ref{DStest} and the dashed circle the $R_{200}$ of A963\_1. Shaded areas in the bottom panel indicate several regions of interest such as the `recent infall', ram-pressure `stripping', and `virialized' regions, discussed in the text. 
\label{radec_col}}
\end{center}
\end{figure*}

In an attempt to map the large-scale structure (LSS) around A963\_1, in the left-hand panel of Fig.~\ref{LSS} we plotted galaxies with SDSS redshifts in the surroundings of the cluster, and in the redshift range $0.184<z<0.224$. Although 
SDSS spectroscopy is limited at this redshift, the plot shows evidence of at least two filamentary structures feeding the cluster from the North-East (NE) and the South-East (SE). A possible third filament can also be seen from the West (W) direction. 
These structures seem to be separated by void or empty regions, and in phase-space (right-hand side of Fig.~\ref{LSS}) they show narrow velocity ($z$) dispersions at all $r$, which supports the idea that they are indeed cosmic filaments. In fact their phase-space distribution is similar to the large-scale filaments found around Virgo \citep[see Figure 3 in][]{Lee2015} and Coma \citep[see Figure 4 in][]{Falco2014}.
It is possible that the groups (plotted as grey open circles in Fig.~\ref{LSS}) were accreted into the cluster from such filamentary structure. 

In addition to the groups currently being accreted into the cluster, the X-ray  temperature map in the inner region of the cluster (made by AF, not shown here) further reveals a past merger/interaction. These events typically leave inhomogeneities in the temperature map at the cluster core, which are visible in A963\_1.

\subsection{Distribution of red, blue and \HI\ -rich galaxies}
\label{subsec:properties}

In the top panel of Fig.~\ref{radec_col} we inspect the spatial distribution of all the cluster galaxies, highlighting their $NUV-R$ colour and \HI\ content in the same manner as in the CMD of Figure \ref{CMD}. 
It is evident that the cluster core, as well as groups host a significant population of passive, gas-poor galaxies. Passive galaxies also extend beyond the imposed boundaries ($R_{200}$) of the groups, most notably in Groups $\alpha$, C-B and $\gamma$, possibly due to group elongation  caused by the cluster potential well, as discussed in Section~\ref{subsec:substructures}.

\begin{figure}
\includegraphics[scale=0.48]{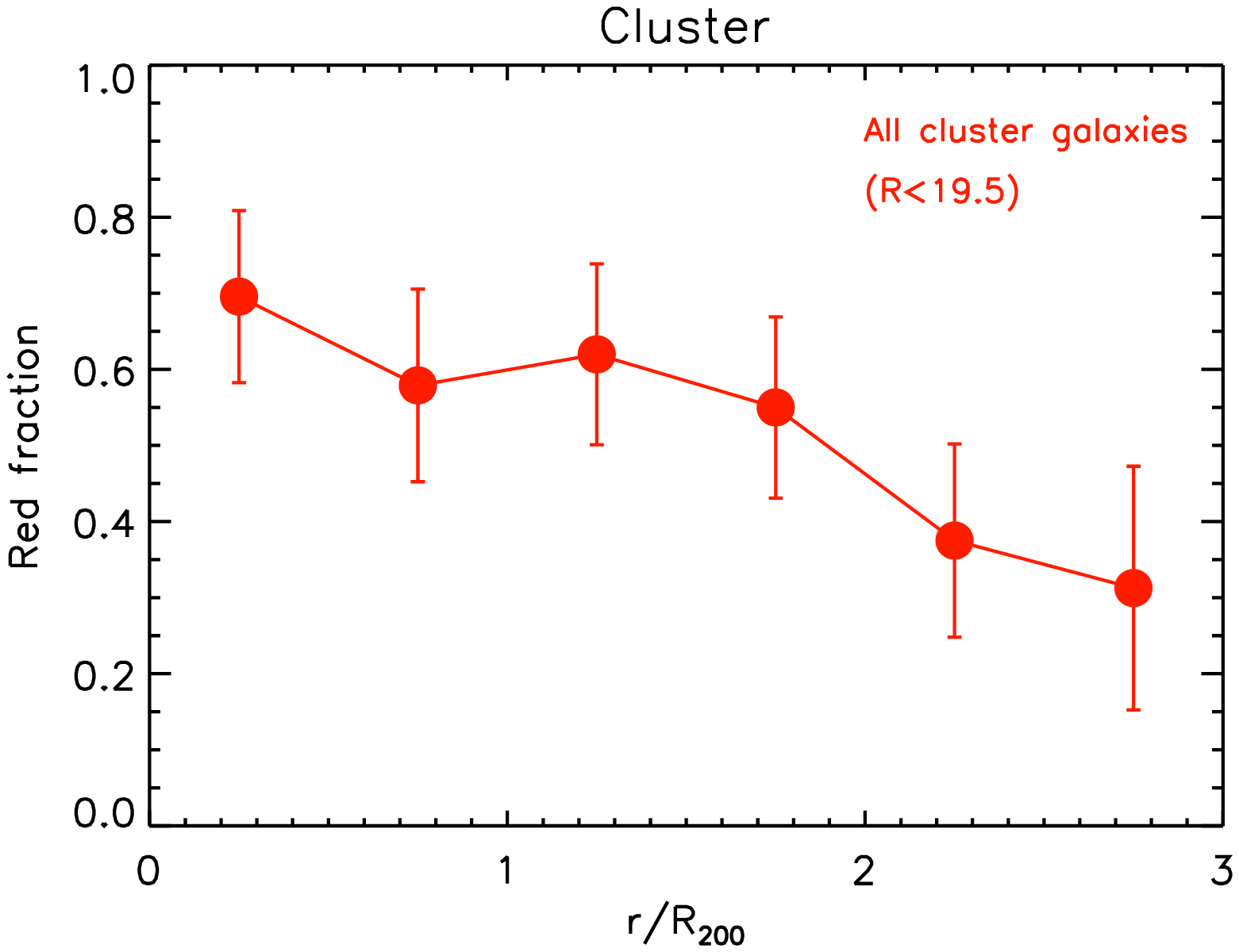} 
\includegraphics[scale=0.48]{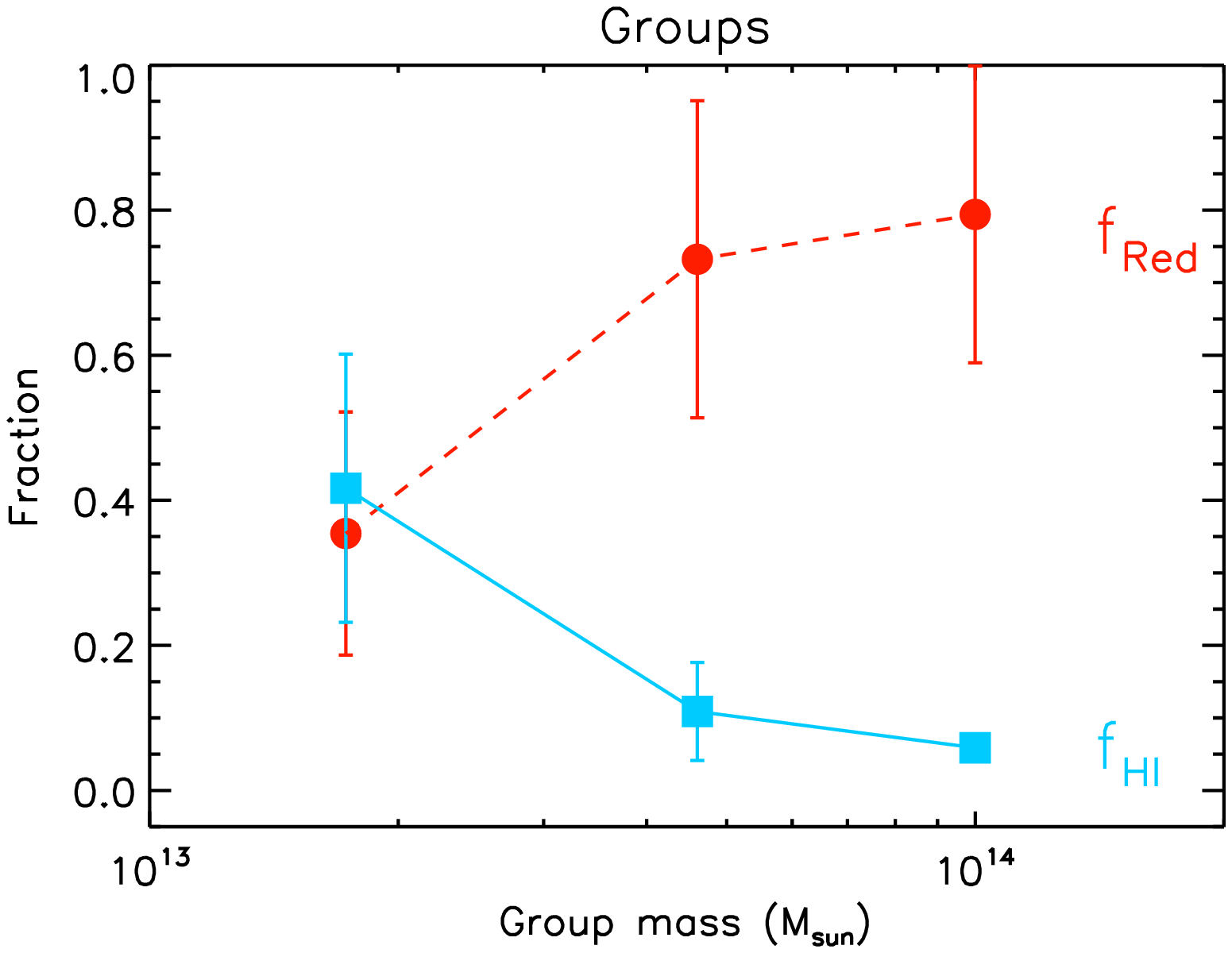}
\caption{Top: The fraction of passive  cluster galaxies brighter than $R=19.5$ as a function of distance from the cluster centre. Bottom: fraction of passive (red circles) and \HI\ -detected ($f_{\HI\ }$ cyan squares) galaxies in the groups A$+\delta$, B$+\alpha$ and C,  as a function of group mass (groups of similar mass were added to increase the number of galaxies and make the Poisson errors smaller). In this case, no magnitude cut was used. 
\label{fracs_r}}
\end{figure}

To study more closely the connection between the passive galaxies and environment, in the top panel of Fig.~\ref{fracs_r} we evaluate the incidence of passive galaxies as function of distance from the cluster centre. 
The plot clearly shows how passive galaxies (which are also typically the most massive ones) dominate the cluster core, and become more scarce with increasing $r$, as expected. 
The bottom panel of Fig.~\ref{fracs_r} further shows, for galaxies in groups with robust mass estimates, how the fraction of passive group galaxies increases with increasing group mass, at the expense of \HI\ -detections. While the fraction of passive galaxies in the lower mass groups coincides with that of the infall region at the largest clustercentric distances ($r>2 \times R_{200}$), at higher group mass, it jumps to $\sim80\%$. 
The observed jump in the passive fraction (as well as a comparable decrease of \HI\ -detections) from low to intermediate-mass groups confirm the existence of a strong connection between the gas content, star formation activity and host halo mass, and suggest that environmental star formation quenching starts being significant in groups more massive than $\sim2\times10^{13}M\odot$.   
We note that for the plot in the bottom panel of Fig.~\ref{fracs_r} we did not make any magnitude cut due to the low number of galaxies. We note however that the stellar mass distribution of the group galaxies are very similar, ranging from $\sim10^9$ to $\sim10^{11} M\odot$ and with mean value $= 3.55 \times 10^{10} M\odot$.

The bottom panel of Fig.~\ref{radec_col} shows the distribution $NUV-R$ colour and \HI\ in projected position versus velocity phase-space. 
A similar plot was already shown in \citet{Jaffe2015} but it is presented here again with the much larger spectroscopic sample. 
A few areas, defined in \citet{Jaffe2015}, are shown.  The `recent infall' and `virialized' areas reflect the orbital histories of cluster galaxies, as already discussed in Section~\ref{subsec:substructures}. The `stripping' zone, also  defined in \citet{Jaffe2015}  is the outcome of embedding \citet{GunnGott1972}'s prescription of RPS, together with a model of A963\_1's ICM, into dark matter cosmological simulations. 
Our model predicted gas-rich galaxies to be located in the `recent infall' region of phase-space, avoiding the `stripping' and `virialized' zones. 
The observed distribution of \HI\ -detections is in striking agreement with the model \citep[see Figure 6 in][]{Jaffe2015}. 
The idea is that as galaxies fall into the cluster from high clustercentric distances (`recent infall' region in phase-space), they can gain velocity as they approach the cluster core, which leads them into the `stripping' zone. After their first pass, they will oscillate in phase-space slowly settling in the potential well of the cluster, building up the `virialized' region. 

Fig.~\ref{radec_col} further shows that passive galaxies dominate both the central part of the cluster, and the `virialized' part of phase-space, suggesting that many of them have been in the cluster for a long time ($\gtrsim 4$~Gyr). 
However, not all passive galaxies are in the virialized region. In fact, a significant population  of passive galaxies can be seen in the infall region of the cluster.
It is possible that some of these (especially the most massive ones) have been `mass quenched'  \citep[by internal processes such as AGN feedback; see][]{Peng2010}, but additional `environmental quenching' might also be at play, especially for lower mass galaxies. 
As discussed in \citet{Jaffe2015}, the abundance of passive galaxies in the cluster outskirts cannot easily be explained by RPS alone (or backsplashing galaxies). 
However, a more plausible explanation for this effect is that many of these galaxies started to be quenched in a previous environment. One likely possibility, given the results shown in Fig.~\ref{fracs_r}, is group `pre-processing'. In fact, $\sim15$\% of the passive galaxies in the infall region of the cluster belong to the groups identified in Section~\ref{subsec:substructures}. This is a lower limit however, as we can only identify the most obvious group members and many group galaxies have probably mixed with the cluster population already. In addition, the limited area of the XMM observations does not allow the identification of X-ray groups at large clustercentric distances.

\begin{figure}
\centering
 \includegraphics[scale=0.48]{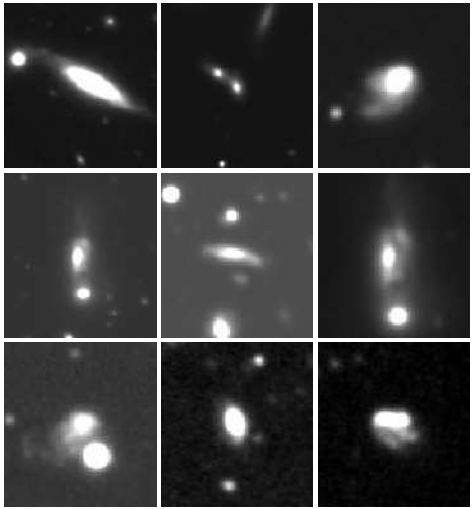}
\caption{Example Subaru $V$-band images ($\sim 20$arcsec$\times20$arcsec) of the peculiar galaxies identified in A963\_1. Most of the peculiar galaxies show signs of interactions with another galaxy, but a few could be stripped by the ICM. The three most convincing `jellyfish' candidates are shown in the bottom row. \label{pec_gals} }
\end{figure}

\subsection{Morphological peculiarities}
\label{subsec:morphologies}

In the quest to detect signs of the different physical processes driving galaxy evolution in this actively accreting cluster, two of the authors of this paper (YJ and MMC) visually inspected the morphologies of the galaxies in the central $\sim30$~arcmin$^2$ region of the cluster using the Subaru $Ic$ and $V$ images. We found many galaxies with peculiar morphologies. 
Most of them show signs of gravitational interactions with another galaxy, such as `bridges' of stars, multiple nuclei, or irregular morphologies, although the origin of the perturbations is not always clear. Some examples are shown in Fig.~\ref{pec_gals}. 
A few of the peculiar galaxies exhibit one-sided `tails' or `jellyfish' morphologies,  typically associated with galaxy-ICM interactions. The most convincing cases are shown in the bottom row of  Fig.~\ref{pec_gals}. 
In our analysis, we treat all peculiar galaxies as `interacting', not attempting to identify the cause of their disturbances. 

\begin{figure}
\centering
 \includegraphics[trim=20 0 20 0,scale=0.49]{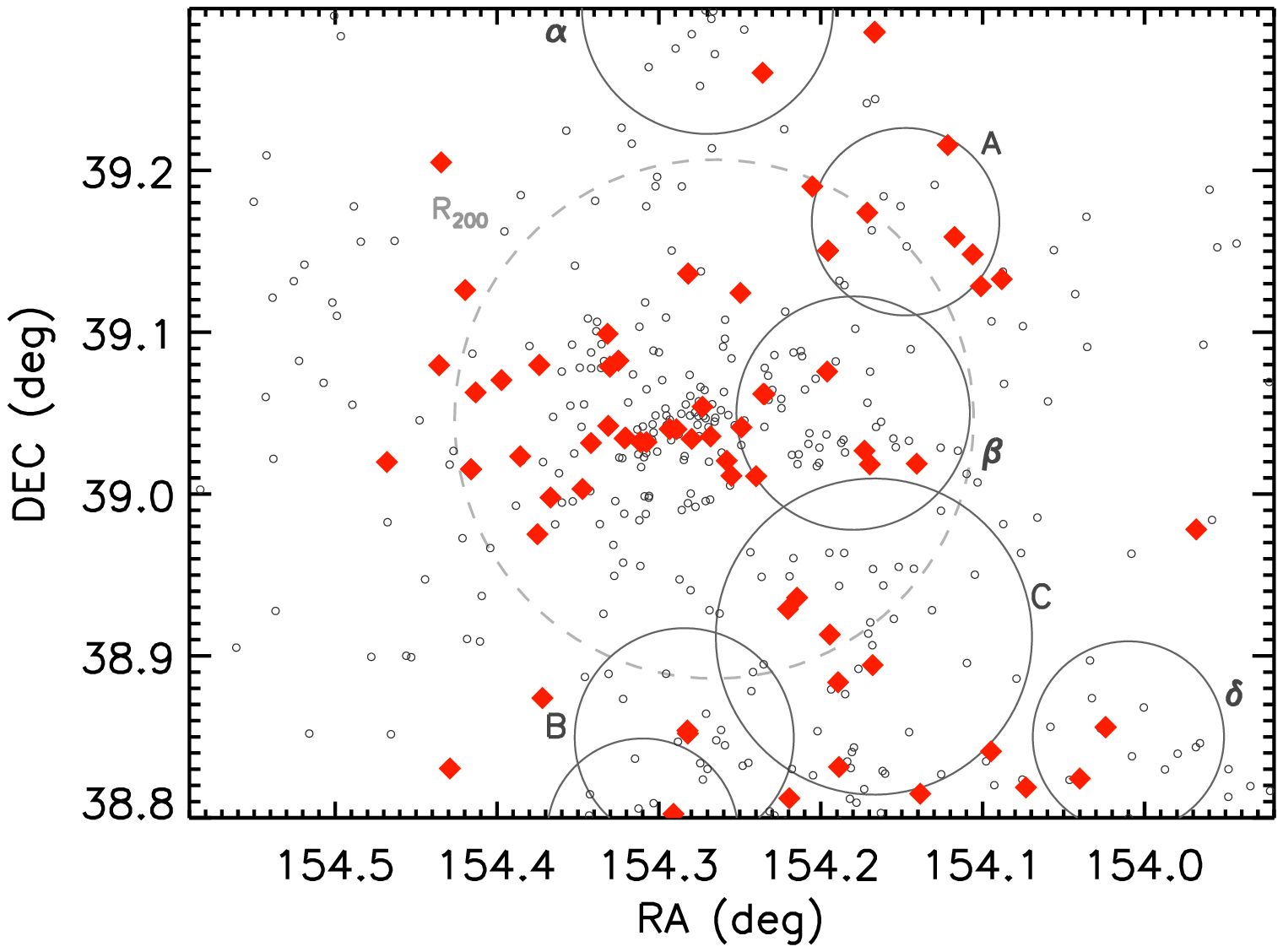}
 \includegraphics[trim=20 0 20 0,scale=0.49]{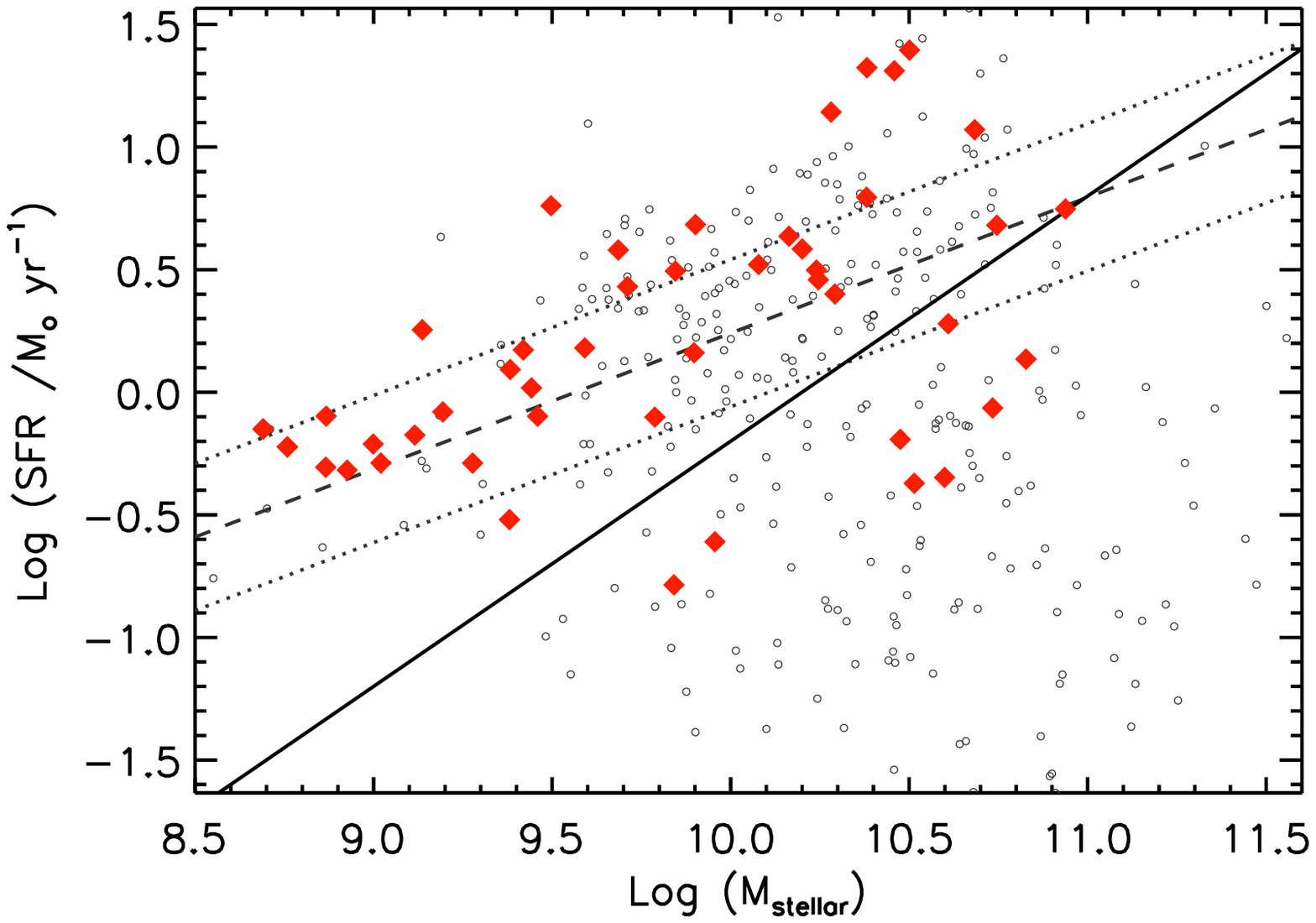}
\caption{The spatial  (top), and SFR versus $M_{*}$ (bottom) distribution of all cluster members (no magnitude cut; open circles) are compared with galaxies with peculiar morphology (filled red diamonds). Note that morphologies are only available in the area covered by Subaru (plotted area in top panel). The bigger open circles in the top panel delimit the different groups as in Fig.~\ref{radec_col}. The dashed circle in the centre corresponds to the $R_{200}$  of the cluster. The lines in the bottom panel are as in Fig.~\ref{sfs}.
\label{Env_eff}} 
\end{figure}

The top panel of Fig.~\ref{Env_eff} shows the distribution of the cluster galaxies in the sky. 
Peculiar galaxies (solid symbols) clearly trace the structures of A963\_1: they cluster near the centre (extending to the east), and they are also common in and around the groups. 
When selecting only group galaxies, we find that peculiar galaxies are more common in the lower-mass groups A, B, and $\delta$ ($46$\% on average, for galaxies with $R<19.5$) in comparison with Group C ($13$\%). Moreover, when inspecting the $NUV-R$ colours of the peculiar group galaxies, we find that  in grup C, they are all passive, while in the lower-mass groups they are a mixed population.

The SFR-stellar mass diagram of the bottom panel of Fig.~\ref{Env_eff}, further shows that the vast majority  ($\sim 80\%$) of peculiar galaxies are actively forming stars. 

The abundance of peculiar star-forming galaxies in groups is not a surprise, since  most peculiar galaxies in our sample are clear mergers, a phenomena more likely to occur in galaxy systems of low velocity dispersion.  
The large number of peculiar galaxies in the central part of the cluster however, is harder to explain. 

Using cosmological simulations \citet{Vijayaraghavan2013} show that during a group's pericentric passage within a cluster, galaxy-galaxy collisions are enhanced. In addition, merger shock enhances the ram-pressure on group and cluster galaxies and an increase in local density during the merger leads to greater galactic tidal truncation. They call the combination of these effects `group post-processing'. 

It is possible that the high fraction of peculiar (and/or star-forming) galaxies near the cluster core arise via post-processing by the numerous groups falling into A963\_1 and the very high galaxy density. This is supported by the high degree of substructure in this cluster, and the presence of not only mergers, but also ram-pressure stripped galaxies. 

\begin{figure}
\centering
 \includegraphics[trim=10 0 0 0, scale=0.56]{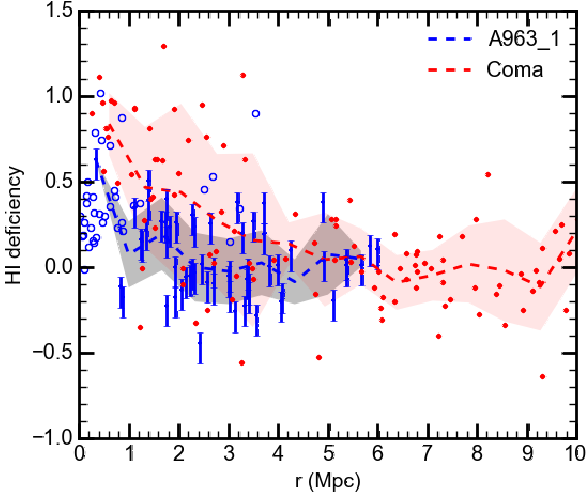}
\caption{\HI\ -deficiencies as a function of distance from the cluster centre for blue galaxies in A963\_1 (blue circles) and the Coma cluster \citep[red triangles;][]{BoselliGavazzi06}. For A963\_1 galaxies, \HI\ deficiencies for \HI\ -detections were computed using the average of all morphological types, with error bars corresponding 
to the \HI\ -deficiencies of spirals from Sa to types later than Sc. For non detections (open circles) we plot lower limits on the \HI\ deficiency. All points were corrected for primary beam attenuation. 
The solid and dashed lines trace the mean \HI\ deficiency values for the galaxies in A963\_1 and Coma respectively, and the shaded regions correspond to $1 \sigma$ uncertainties.
 \label{HIdef}}
\end{figure}

\begin{figure}
\centering
 \includegraphics[trim=20 0 10 0, scale=0.67]{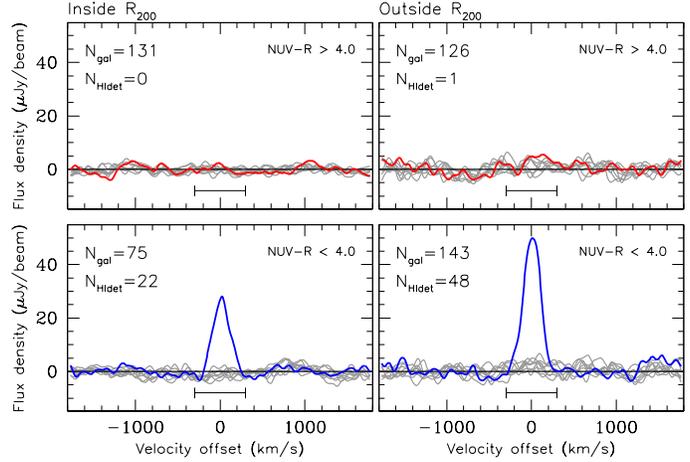}
\caption{\HI\ stacking of passive (top) and blue (bottom) cluster galaxies (excluding group members) inside (left) and outside (right) $r=R_{200}$. These stacks exclude group galaxies. In each panel, the thin grey lines  correspond to the eight reference offset stacks. The number of galaxies that went into each stack ($N_{\rm gal}$) and number of direct \HI\ -detections ($N_{\HI\ {\rm det}}$) are indicated in the top-left corner of each panel. The horizontal  bar underneath the spectra indicates the rest-frame velocity range used in the flux measurement. 
 \label{stack_R200}}
\end{figure}

\subsection{The \HI\ content of galaxies as a function of environment}
\label{subsec:HIstacks}

\subsubsection{\HI\ deficiencies}

We first assess the \HI\ content of galaxies in A963\_1 by computing \HI\ deficiencies. 
From a sample of 324 (mostly spiral) isolated galaxies \citet{HaynesGiovanelli1984} defined \HI\ deficiencies as the logarithmic difference
between the observed \HI\ mass and the expected value in isolated objects of similar morphological type and linear size.  
%

Unfortunately, the visual classification of A963\_1 galaxies from available ground-based images (see Section~\ref{subsec:morphologies}) could only allow the identification of interacting galaxies but not a detail assessment of T-type morphologies. 
For this reason, we assumed all blue galaxies are spirals and all passives ones are elliptical or S0, and computed \HI\ deficiencies for all the blue ($NUV-R<4$) galaxies in A963\_1.   
Galaxy sizes were determined from the 50\% light radii in B-band and the angular distance to A963\_1. As a comparison sample for the computation of the deficiencies we used the local field galaxies of \citet{HaynesGiovanelli1984}. 
For galaxies not directly detected in \HI\ we compute the lower limit of \HI\ -deficiency correcting for primary beam attenuation.  

Fig.~\ref{HIdef} shows the computed  \HI\ -deficiencies as a function of clustercentric distance. The plotted error bars  correspond to the \HI\ -deficiencies of  Sa spirals to types later than Sc. 
In addition, we plot Coma galaxies from \citet{BoselliGavazzi06} as local reference. Note that A963\_1 and Coma are similarly massive ($\sim 10^{15} M\odot$) clusters. 
Despite the scatter, the figure clearly shows how galaxies in both clusters become increasingly more deficient as they get closer to the cluster centre. 
The radial decay of \HI\ deficiencies in A963\_1 agrees within errors with the trend observed in Coma, although close to the cluster core, the z=0.2 galaxies are a factor of 2 less deficient than the local ones.

\subsubsection{\HI\ stacking in and around the cluster}
As in \citet{Verheijen2007}, we also probe deeper into the gas content of the cluster galaxies by stacking the \HI\ spectra of galaxies with known optical redshifts in bins of colour and environment. To do this, we extract the spectra around the expected location of the 21~cm line in each individual galaxy using its \HI\ position and redshift when available, or the optical information otherwise. We do this over an area of roughly the size of the FWHM of the synthesized beam, and  401 channels around the galaxy, corresponding to a window of $\sim3600$ km~s$^{-1}$ in velocity. 
This generous selection allows us to quantify the noise around the 21~cm line robustly. 
We then add all the spectra in a given sample of galaxies to compute an average flux, that is later converted into a $M_{\HI\ }$ assuming all galaxies are at the mean distance of the cluster ($z_{\rm cl}$):   
  $M_{\HI\ } = 2.36\times10^5 \times d^2 \times S{\rm d}v$, 
where $d = 1014$~Mpc (the distance to A963\_1), $S{\rm d}v$ = $\sum  S_i\times {\rm d}v \times(1+z_{\rm cl})$,  $S_i$ the flux density in channel $i$, and ${\rm d}v=8.14$ km~s$^{-1}$, the rest-frame channel width.

To estimate the errors on the \HI\ fluxes (and consequently $M_{\HI\ }$), for each stack, eight reference stacks at different offset positions were made. We then calculated the variance in the eight reference fluxes, integrated over the same velocity windows as the on-target stack. The $M_{\HI\ }$ values derived from the stacks presented in this section, and their associate errors are listed in Table~\ref{HIstack_table}.

\begin{figure*}
\centering
 \includegraphics[trim=0 0 0 0, scale=0.72]{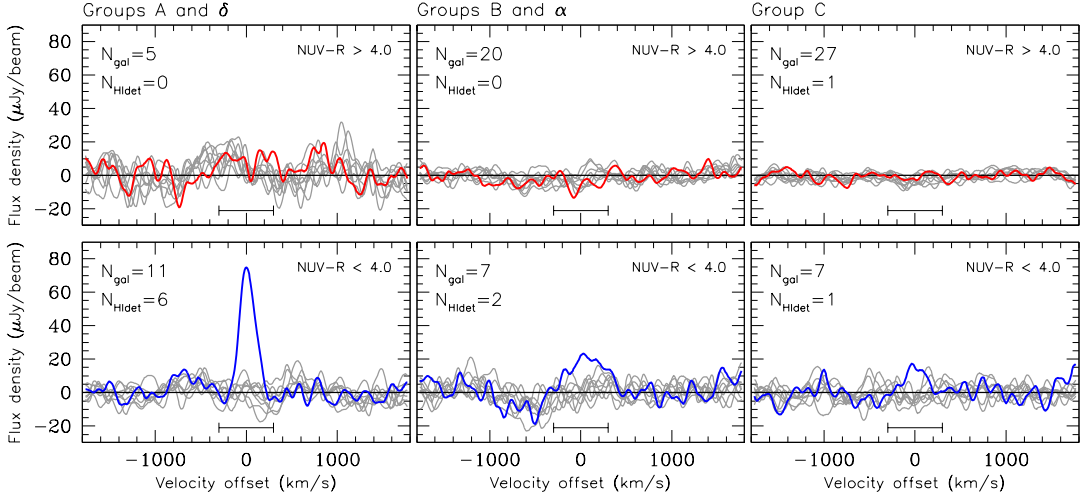}
\caption{\HI\ -stacking of red (top row) and blue (bottom) galaxies in groups, separated in three bins of group mass: low (left, Groups A and $\delta$), intermediate (middle, groups B and $\alpha$), and high mass (right, group C).   No magnitude cut was applied. Lines and labels are as in Fig.~\ref{stack_R200}.  
 \label{HIstack_groups}}
\end{figure*}

\begin{figure*}
\centering
\includegraphics[trim=0 0 0 0,scale=0.72]{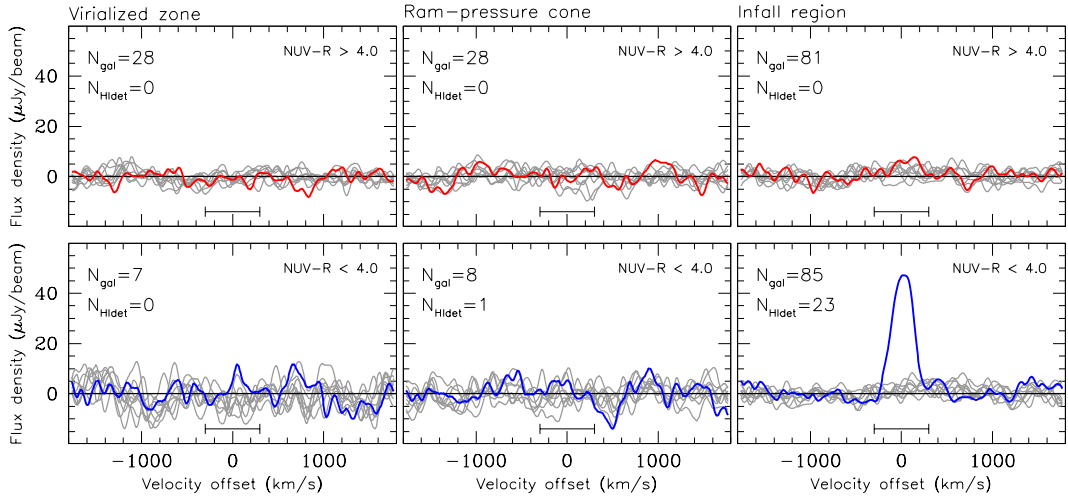}
\caption{\HI\ -stacking of passive (top) and blue (bottom) galaxies in the `virialized' (left), ``ram-pressure'' (middle) and `recent infall' (right) regions of phase-space, as defined in the bottom panel of Fig.~\ref{radec_col}. Lines and labels are as in Fig.~\ref{stack_R200}. Only galaxies above the spectroscopic completeness limit ($R<19.5$) were considered. 
\label{stack_pps}}
\end{figure*}

\begin{figure*}
\centering
  \includegraphics[trim=60 0 0 0, scale=0.66]{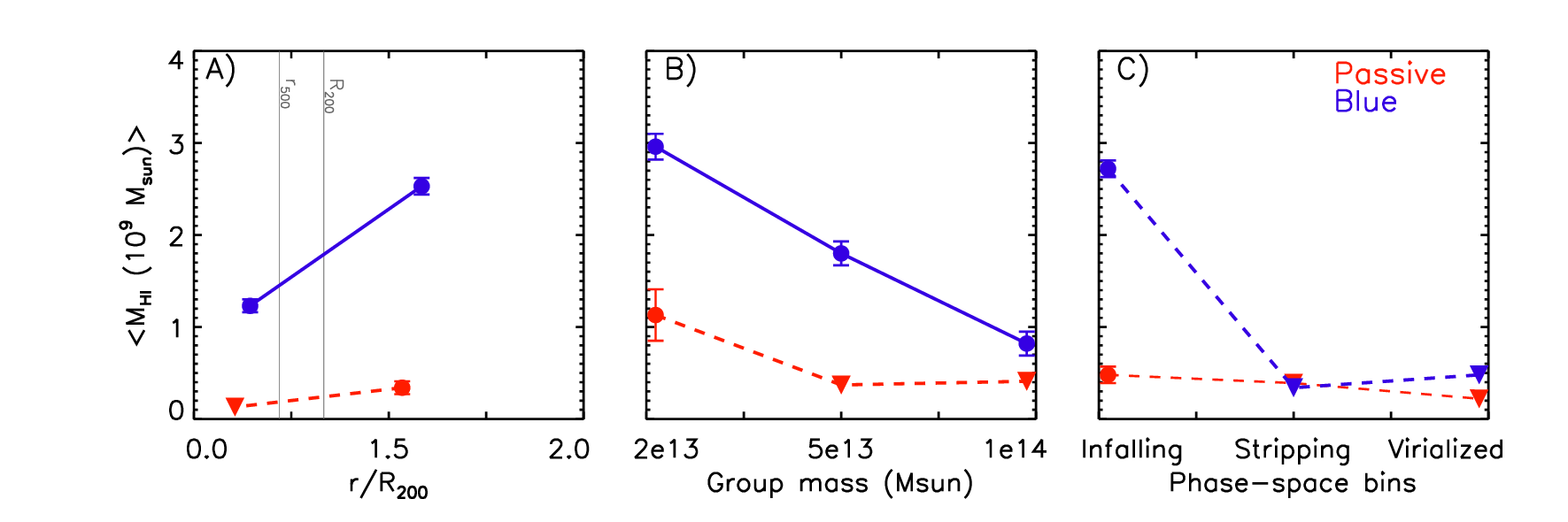}
\caption{Results from the \HI\ stacking in different environments. \textbf{(A)} The average \HI\ mass of passive and blue galaxies inside and outside $R_{200}$ is compared for all cluster members 
with $R<19.5$, excluding galaxies in groups. We plot the mean $r/R_{200}$ value in each bin against the average $M_{\HI\ }$. 
\textbf{(B)} The average $M_{\HI\ }$ in groups of similar mass (see Fig.~\ref{HIstack_groups}) is plotted against average group mass for passive and blue galaxies. 
\textbf{(C)} Average $M_{M\HI\ }$ in different regions of phase-space for passive and blue ($R<19.5$) galaxies. In all panels solid lines and solid circles are used when there was a measured $M_{\HI\ }$ whereas dashed lines and inverse triangles are used when only upper limits could be measured.
 \label{HIstacks_plots}}
\end{figure*}

We demonstrate the power of stacking in Appendix~\ref{sec:app_stacks}. The top panel of Fig.~\ref{HIstack_CMD} shows the results of the \HI\ stacking in different bins of colour and magnitude. It is noticeable how the 21~cm emission decreases dramatically with redder colour and fainter magnitudes, as expected. To further test whether the direct \HI\ detections dominate the stacks, we performed the same stacks in colour and magnitude bins, but excluding the direct \HI\ detections (bottom panel of Fig.~\ref{HIstack_CMD}). The trends seen before stand,  although naturally, the total \HI\ masses in this case decrease. However, the bulk of the galaxies not directly detected in \HI\ do have significant amounts of gas, in many cases not much below the threshold for the detected ones. The results of this exercise are summarized in Fig.~\ref{HIstack_CMD_plot}. 

In the following, we present several \HI\ stacks as a function of environment, considering \HI\ -detected and non-detected galaxies combined.  We use bins of colour, which is a proxy for star formation activity, and only consider galaxies  above our spectroscopic completeness limit ($R>19.5$) unless otherwise stated. We prefer to bin in observable quantities, to avoid contamination from poor modelling.

\subsubsection{\HI\ stacking in the substructures}
\label{subsubsec:HIss}

In  Fig.~\ref{stack_R200} we compare the \HI\ content of cluster galaxies in the innermost region of the cluster ($r<R_{200}$) with those in the outskirts of the cluster ($r>R_{200}$). We do this using the spectroscopically complete sample in the colour bins defined in Fig.~\ref{CMD} (i.e. blue and red/passive), and excluding galaxies in groups. 
The stacks show a mild detection of \HI\ emission ($\langle M_{\HI\ } \rangle = 0.34 \times 10^9 M\odot$)in the passive galaxies residing in the outskirts of the cluster, and no sign of emission ($M_{\HI\ } < 0.13 \times 10^9 M\odot$) in the innermost part. In contrast, when stacking blue cluster galaxies a prominent emission is seen. At $r>R_{200}$, $\langle M_{\HI\ } \rangle =2.53 \times 10^9 M\odot$, while in the cluster core $M_{\HI\ }$ is reduced by $\sim 50\%$. 
The average fluxes and $M_{\HI\ }$ of the stacks are listed in Table~\ref{HIstack_table} and the results are summarized in panel A of Fig.~\ref{HIstacks_plots}, where the solid blue line shows the reduction of \HI\ towards the cluster centre. 

In Section~\ref{subsec:substructures} we identified seven groups falling into the cluster, that show an increasing fraction of passive galaxies with increasing group mass at the expense of \HI\ -detections (Fig.~\ref{fracs_r}).  To further investigate the \HI\ content of the groups, in Fig.~\ref{HIstack_groups}, we stack the \HI\ emission of group galaxies in bins of colour and group mass.  We only considered groups with reliable mass estimates, and  merged together galaxies in groups of similar mass: A and $\delta$ (average mass of  $1.75\times 10^{13} M\odot$); B and $\alpha$ ($4.6 \times 10^{13} M\odot$), and C ($1.0 \times 10^{14} M\odot$). 
Due to the reduced number of galaxies in these stacks we did not limit the group samples by magnitude. 

Passive group galaxies (top row in Fig.~\ref{HIstack_groups}) barely host any \HI\. Only passive galaxies in the lowest mass groups have a detection ($\langle M_{\HI\ }\rangle=1.13 \pm 0.18 \times 10^9 M\odot$). Blue group galaxies instead have larger \HI\ gas reservoirs. This can be seen in panel B of Fig.~\ref{HIstacks_plots}. The \HI\ gas masses of blue galaxies in the two lowest mass 
groups ($\langle M_{\HI\ }\rangle=2.96 \pm 0.14 \times 10^9 M\odot$) are consistent with those of non-group blue galaxies in the infall region of the cluster (left-most blue point in panel C.  The \HI\ masses of blue group galaxies progressively declines with group mass, being a factor 3 lower in the highest mass group ($\langle M_{\HI\ }\rangle=0.82 \pm 0.13 \times 10^9 M\odot$).

\subsubsection{\HI\ stacking in phase-space bins}
\label{subsubsec:HIpps}

Fig.~\ref{stack_pps} shows the \HI\ stacks of passive and blue galaxies in the `recent infall',  (ram-pressure) `stripping'  and `virialized'  regions of phase-space. These regions (in that order), roughly trace the journey of a galaxy entering a cluster (i.e. its orbital history). For these stacks we used all cluster members in the magnitude-limited sample ($R>19.5$) not associated with groups.  
The most significant \HI\ detection is that of blue galaxies in the infall region of the cluster (bottom-right panel; $\langle M_{\HI\ }\rangle=2.72 \pm 0.09 \times 10^9 M\odot$). The middle and left-hand panels in the bottom row show how blue galaxies lose their \HI\ reservoirs as they move from the infall region of the cluster inwards.  In fact, $> 90 \%$ of their \HI\ gas is removed on first infall as they cross the RPS region of phase-space. 

For passive galaxies, we only detect \HI\ emission in the infall region of the cluster. This is consistent with \citep{Serra2012} that found that $\sim40$\%  of all early-type galaxies outside the Virgo cluster have \HI\ (with $M_{\HI\ } \gtrsim 10^7 M\_{\odot}$), while in the cluster centre, only $\sim10$\%  of them show \HI\ emission. Moreover, the mild \HI\ emission we measure in passive infalling galaxies ($\langle M_{\HI\ }\rangle=0.48 \pm 0.09 \times 10^9 M\odot$) is not far from the local \HI\ mass function's $M*$, which \citet{Serra2012} measured to be $M*=2\times10^9 M\odot$ in Virgo.

The small amounts of \HI\ gas that passive galaxies may hold before entering the cluster is also lost via ram-pressure to the point of no-detection in the `stripping'  or `virialized'  region of phase-space ($M_{\HI\ } < 0.39 \times 10^9 M\odot$ and $M_{\HI\ }<0.22 \times 10^9 M\odot$ respectively). 
The outcome of these stacks is summarized in panel C of Fig.~\ref{HIstacks_plots}, where it is evident that, although passive galaxies in the infall region of the cluster have \HI\ emission, they have 80\% less \HI\ gas than blue ones. In other regions of phase-space passive galaxies don't show \HI\ emission. This suggests that they could also experience RPS as they move in phase-space into the `stripping' and `virialized' regions.


\subsection{Star-formation activity as a function of environment: stacking optical spectra}
\label{subsec:Opt_stacks}

To understand the effect of gas stripping on the star formation activity of the cluster galaxies, in this section we complement the \HI\ stacking presented in Section~\ref{subsec:HIstacks} with stacking of the optical spectra. 
For homogeneity, we only used the sub-sample of galaxies observed with the MMT, shifting all the spectra to rest-frame, and normalizing the fluxes so that of the continuum between $4400 \rm \AA$ and $5800 \rm \AA$ overlap on average.

\begin{figure}
\includegraphics[trim=30 0 0 0, scale=0.21]{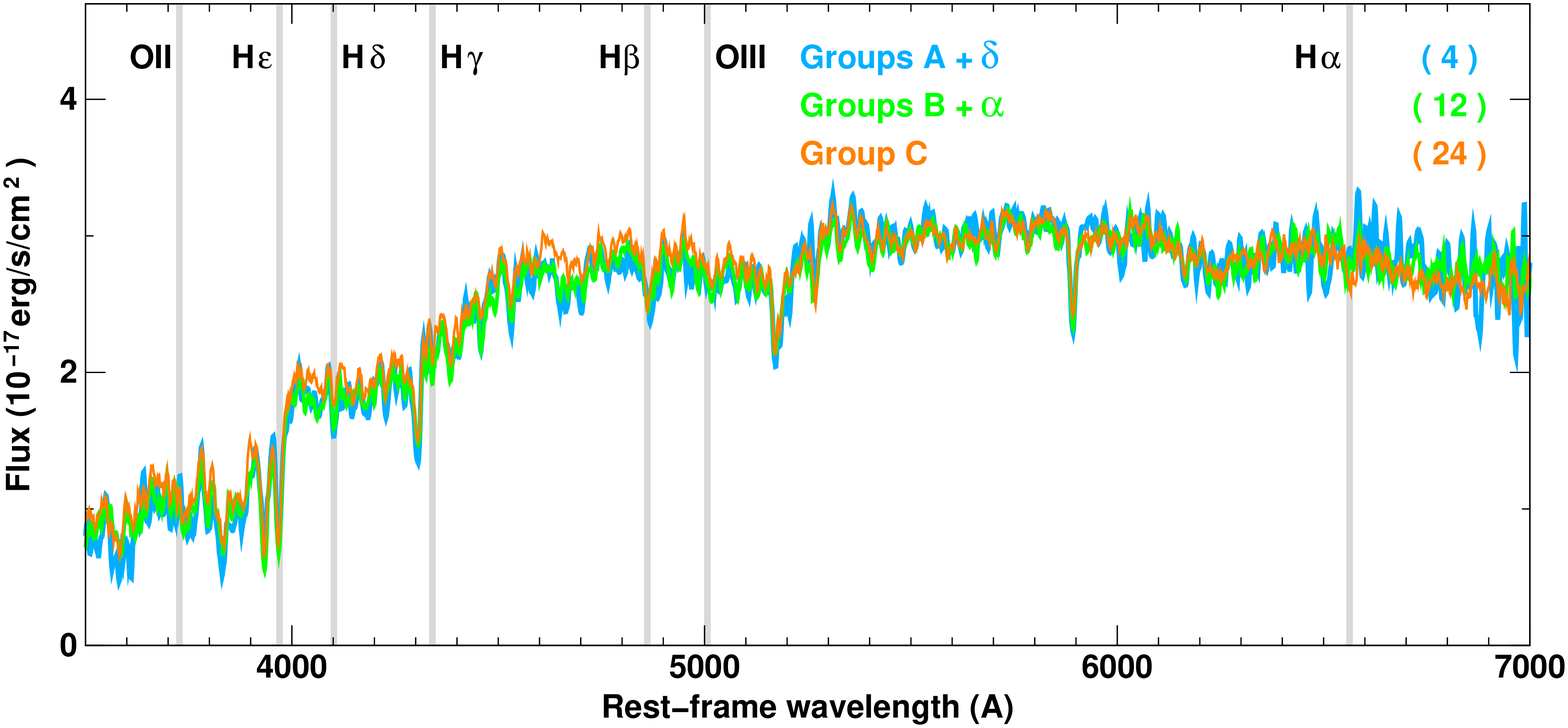}
\includegraphics[trim=30 0 0 0, scale=0.21]{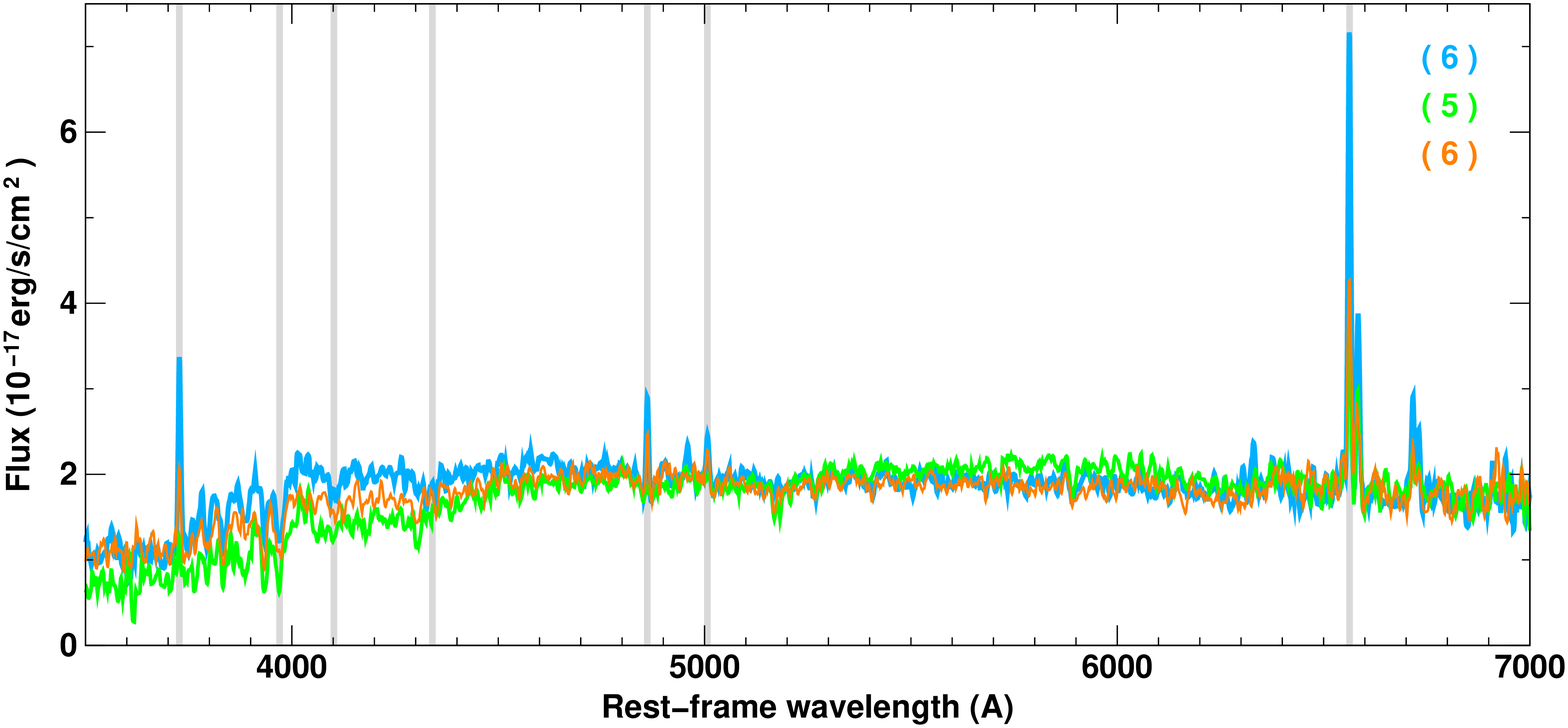}
\caption{Optical stacks of passive (top) and blue (bottom) galaxies inside low (Groups $A+\delta$; blue) intermediate (Groups $B+\alpha$; green) and high mass (Group $C$; red) groups, as labelled. The number of galaxies in each stack is indicated inside the parenthesis in the top-right corner. Main spectral features are also indicated. \label{opt_gr}}
\end{figure}

Fig.~\ref{opt_gr} shows the optical stacks for galaxies in groups, binned by galaxy colour and group mass, as in Fig.~\ref{HIstack_groups}. Again, we did not make magnitude cuts due to the low number of galaxies.   Regardless of the group mass, passive group galaxies (top) show spectra dominated by old stellar populations, with strong absorption features. Conversely, the stacks of blue group galaxies (bottom) display some differences with varying group mass. 
Most notably, the stacked spectra of the blue galaxies in the lower mass groups (A$+\delta$) show much stronger emission lines (EW($H\alpha$)$=27.8 \pm 6.6 {\rm \AA}$, see Table~\ref{OPTstack_table}) and a blue continuum, than the blue galaxies in the intermediate-mass groups B$+\alpha$ or in the most massive group C.

To test the strength of the stacking results presented in the bottom panel of Fig.~\ref{opt_gr} (i.e. stacking of blue group galaxies, see also Table~\ref{OPTstack_table}), we stacked five randomly selected blue galaxies from groups A$+\delta$ and obtained an EW($H\alpha$) equal or lower than that measured in groups B$+\alpha$ only $1.9\%$ of the time. Moreover, we could never reproduce the EW($H\alpha$) of groups A$+\delta$ from the members of group B$+\alpha$ as non of their individual EW($H\alpha$) as as high as $27.8{\rm \AA}$.

These results indicate that low-mass groups still hold a significant population of star-forming galaxies, while groups more massive than $\sim 2\times 10^{13} M\odot$ already show signs of star formation quenching.

\begin{figure}
\includegraphics[trim=30 0 0 0,scale=0.21]{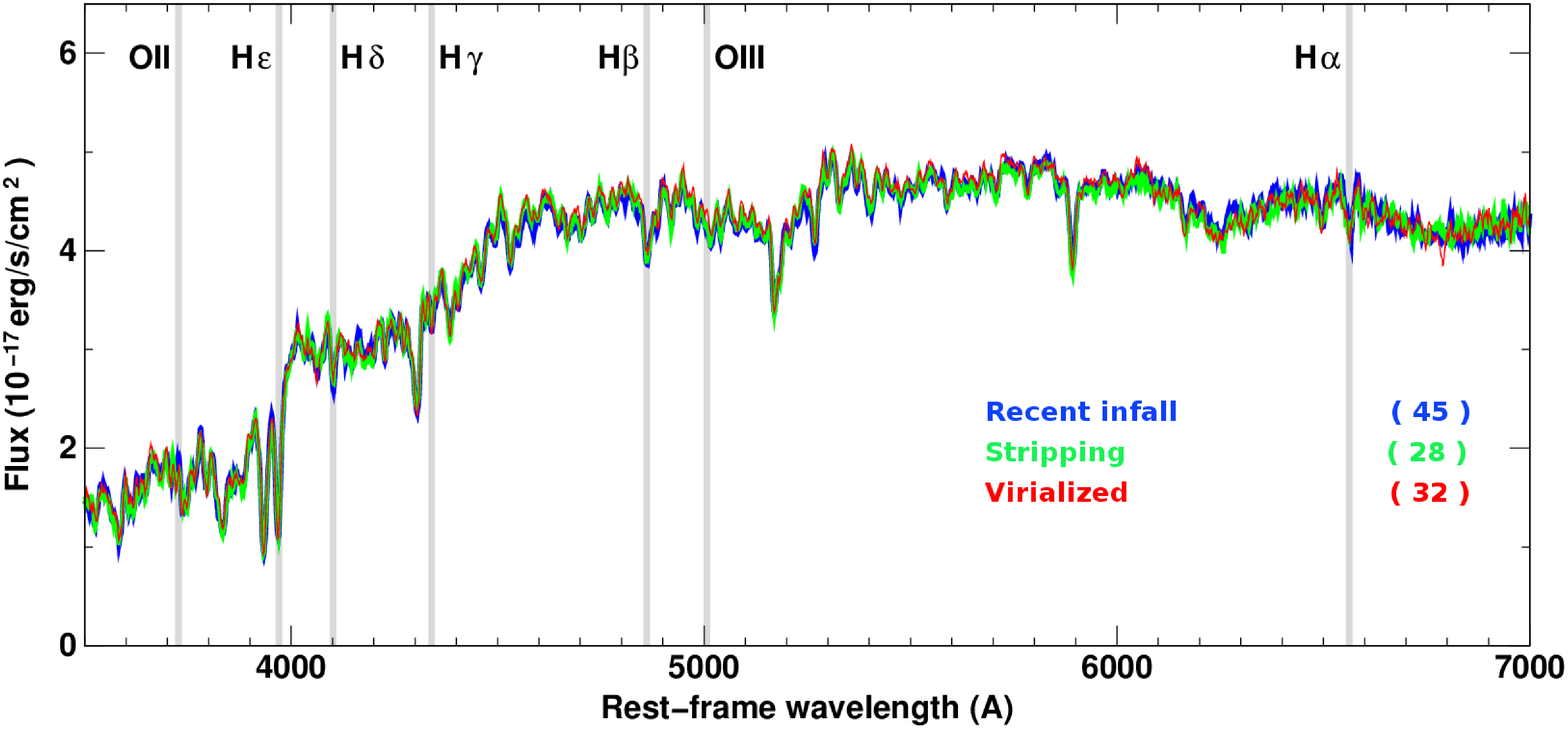}
\includegraphics[trim=30 0 0 0,scale=0.21]{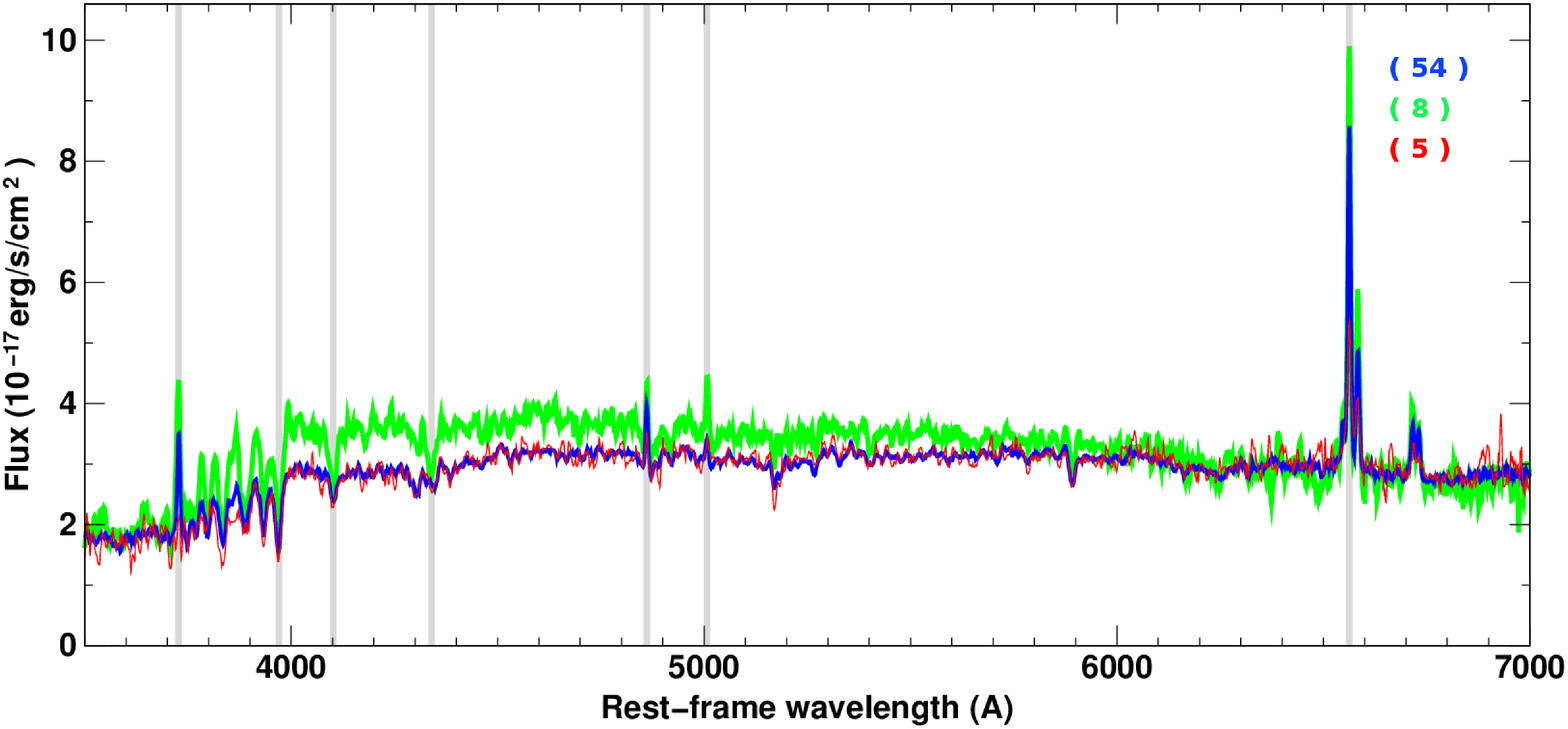}
\caption{Optical stacks of passive (top)  and blue (bottom) galaxies brighter than  $R=19.5$ in different regions of phase-space:  the `recent infall' (blue curves), `stripping' (green) and `virialized' regions (red), as labelled.  As in Fig.~\ref{opt_gr}, the main absorption and emission-lines are marked, and the number of galaxies in each stack is indicated. \label{opt_pps}}
\end{figure} 

\begin{table}
\begin{tabular}{lcc}
\hline
Optical stacking sample	& No.  galaxies	&  EW($H\alpha$)		\\
\hline
\textbf{By group mass:}			&		&				\\
Groups A$+\delta$			& 6		& $27.8 \pm 6.6$  		\\
Groups B$+\alpha$ 			& 5		& $9.4 \pm 3.8$ 			\\
Group C					& 6		& $14.7 \pm 5.1$			\\
\hline
\textbf{By location in phase-space:}	& 		& 				\\
Recent infalls				& 54		& $18.1 \pm1.7$			\\
Stripping 				& 8		& $21.5 \pm 8.8$			\\
Virialized				& 5		& $9.6 \pm 1.8$			\\
\hline
\end{tabular}
\caption{This table lists the results from the optical spectra stacks performed to blue (star-forming) galaxies in different regions of the cluster, namely groups of different mass and location in phase-space, as indicated in the fist column. The second column indicates the number of galaxies used in each stack and the third the measured EW($H\alpha$) with its bootstrapping error.}
\label{OPTstack_table}
\end{table}

In Fig.~\ref{opt_pps} we show the stacked spectra of galaxies in the `recent infall' (blue spectrum),   `stripping' (green) and `virialized' (red) regions of phase-space. As for the \HI\ stacks of Fig.~\ref{stack_pps}, we used all cluster members in the magnitude-limited sample ($R>19.5$) not associated with groups.  When separating red passive galaxies from blue star-forming  ones, we find that,  the passive galaxies in the different phase-space bins have a very similar appearance, typical of old stellar populations. 
Star-forming galaxies display a much larger variation. The stacked spectra of star-forming galaxies in the `stripping' zone shows a significantly bluer continuum and mildly stronger emission lines  (EW(H$\alpha$)=21.5 $\pm 8.8 \rm \AA$) than the stack of galaxies in the `recent infall'  region (although within errors, see Table~\ref{OPTstack_table}). This difference is bigger  for the `virialized' region (EW(H$\alpha$)=9.6 $\pm 1.8 \rm \AA$).

We computed the likelihood of getting a stacked spectrum with an   EW(H$\alpha$) as low or lower than the one measured in the `virialized' region by stacking at random,  five blue galaxies from the `recent infall'  region. This exercise yielded a probability of $1.86\%$ of producing a stack with EW(H$\alpha$)$\leq9.6 \rm \AA$. In addition, we tested the reality of the strong blue continuum seen in the blue galaxies in the `stripping' zone by looking at their individual spectra. We find that  $5/8$  $(62.5\%)$ show this feature, while only 10 out of 54 ($18.5\%$) blue galaxies in the `recent infall'  region do.

Our results suggest that there is a brief starburst associated to the first passage through the RPS zone in phase-space, that leads to a latter quenching of their star formation. Such an enhancement in the star formation of  galaxies falling into clusters has been reported in the literature \citep[e.g.][]{GavazziJaffe1985,Porter2008,Mahajan2012}.

Finally, although not shown here, we also stacked the optical spectra of star-forming galaxies ($R<19.5$, $sSFR > 10^{-10.2} {\rm yr}^{-1}$) inside and outside the inner part of the cluster ($r500$), split into those individually detected in \HI\ and those  
with no \HI\ detection. In the cluster outskirts, the stacked spectra of the \HI\ -detected and \HI\ non-detections have indistinguishable continua and emission line strengths (both have EW($H\alpha$)$=18 {\rm \AA}$). 
In the inner region of the cluster ($r<r_{500}$) the stacked spectra of the three star-forming galaxies detected in \HI\ still shows similarly strong emission (EW($H\alpha$)$=21 {\rm \AA}$) and a blue continuum. However, the stack of the remaining 21 star-forming cluster galaxies not detected individually in \HI\ shows a redder continuum and weaker lines (EW($H\alpha$)$=10.6 {\rm \AA}$) a factor of 2 weaker than their counterparts in the cluster outskirts. This suggests that the bulk of star-forming galaxies within $r_{500}$ 
are currently in the process of having their star formation quenched, indicative of a slow quenching process \citep[see e.g.][for a discussion]{Haines2013}, and that this is occurring only after their \HI\ gas reservoirs have been significantly depleted.

\section{Summary and discussion}
\label{sec:conclu}

We combine HI, IR, optical, UV, and X-ray data from BUDHIES, SDSS, and LoCuSS in and around A963\_1, a massive cluster at $z=0.2$, to study the effect of (current and past) environment on the transformation of galaxies. In this section, our main findings are first summarized and then discussed. 

\begin{itemize}

 \item We combine \textit{XMM-Newton} observations and hundreds of new MMT redshifts from LoCuSS with BUDHIES spectroscopic and \HI\ data to better characterize A963\_1. The X-ray data, together with a dynamical analysis of the cluster galaxies, reveal that A963\_1 is a $10^{15}M\odot$ cluster with a large amount of substructure. We identify seven groups  with masses between $10^{13}$ and $10^{14} M\odot$ at $\sim 0.5-1.5\times R_{200}$. We further use SDSS data to study the large scale structure around the cluster (beyond $3 \times R_{200}$), and find that at least two filamentary structures are feeding the cluster. We speculate that the groups within the cluster were accreted via such filaments.  

 \item When examining the properties of the \HI\ -detected galaxies in  A963\_1, we find that they have blue colour ($NUV-R<$4), sSFRs typically above $10^{-10.2} {\rm yr}^{-1}$, and are preferentially located outside the cluster core. Passive galaxies on the other hand, dominate the cluster core and become increasingly less frequent at larger clustercentric radii. However, the incidence of passive galaxies outside $R_{200}$ is significant. 

 \item Groups host at least $15\%$ of the passive galaxies in the outskirts of the cluster. Groups more massive than $\sim2\times10^{13}M\odot$ have $\sim 80\%$ of their galaxies quenched, and lower mass groups only $\sim 35\%$. The increase of passive galaxies with group mass happens at the expense of \HI\ detections. 
 
 \item From a position versus velocity phase-space diagram we find that: (i) the infall region of the cluster (`recent infall' zone) contains a mixed population of passive, star-forming and \HI\ -detected galaxies; (ii) the `stripping' zone, where RPS is most intense,  hosts passive galaxies and a few star-forming galaxies not detected in HI; (iii) the `virialized' zone, where the galaxies that have been in the cluster for over a cluster-crossing time are, is heavily populated with passive galaxies not detected in HI.  
 
 \item  We use Subaru $Ic$ and $V$-band images from LoCuSS to visually inspect the morphologies of the galaxies in the cluster and its sub-groups. We find many peculiar galaxies, with signatures of galaxy-galaxy interactions, and in very few cases, `jellyfish' features indicative of intense RPS. Peculiar galaxies are common in and around the low-mass groups indentified inside the cluster, while less common (but still present) in the most massive group. Interestingly, peculiar galaxies are also found in the central region of the cluster. The bulk of peculiar galaxies have enhanced star formation. 

 \item We compute the mean \HI\ deficiency of blue galaxies as a function of distance from the cluster centre and find an increased deficiency (by a factor of $\sim1.5$) in the cluster core. The radial decay of \HI\  deficiencies
in A963\_1 agrees within errors with the observations of the (nearby) Coma cluster, although close to the cluster core the $z = 0.2$ cluster galaxies are a factor of 2 less HI deficient than those in Coma. 
 
 \item We stack the \HI\ emission in different regions of the cluster, and find that: 
 (i) Galaxies in the cluster core have significantly lower $M_{\HI\ }$ than those in the outskirts; (ii) the \HI\ content of group galaxies drops significantly from $\sim1\times 10^{13}M\odot$  to $\sim1\times 10^{14}M\odot$ groups; (iii) when stacking in phase-space, we find that blue star-forming galaxies (likely falling into the cluster from the field), have their \HI\ dramatically reduced on their first passage trough the ICM (i.e. the `stripping' zone in phase-space).

 \item We also stack the optical spectra in different regions of the cluster. The stacked spectra of group galaxies shows that low-mass groups still host significant star formation activity, while higher mass groups are dominated by passive galaxies. Moreover, we look at the stacked spectra of galaxies in different phase-space regions and find mild evidence for a star formation enhancement in galaxies experiencing RPS as they pass though the ICM for the first time.  Finally, we find signs of suppression of star formation in the cluster core when stacking star-forming \HI\ -poor galaxies.
 
\end{itemize}

Our results reveal a complex picture in which several physical mechanisms are responsible for the transformation of galaxies in clusters that are growing through the accretion of  groups. We discuss the most relevant in the following.  
\newline

\textit{Ram-pressure stripping: } The phase-space distribution and \HI\ gas content of blue (star-forming galaxies) in the cluster strongly suggests that RPS by the ICM plays an important role in gas removal. 
In fact, \HI\ and optical stacking of star-forming galaxies in phase-space revealed that the galaxies' \HI\ reservoirs are dramatically reduced as they first pass through the cluster, possibly causing a temporary burst of star formation. Subsequently their star formation activity decreases and eventually halts as they settle into the cluster. Our results are consistent with simulations, that show that as ($\sim 10^{12} M\odot$) galaxies experience RPS at pericentric passages in massive clusters, their star formation can be moderately enhanced \citep{Bekki2014}. 

To fully understand the link between gas removal and star formation activity, in a future paper (Cybulski et al., in preparation), we will investigate the relative efficiency of CO versus \HI\ stripping in  A963\_1 using data from COOL BUDHIES \citep[][]{Cybulski2015}. 
\newline

\textit{Group pre-processing: } 
Although, RPS is clearly affecting young infalling galaxies, the large fraction of passive galaxies in the infall region  of the cluster suggest that RPS is not the only mechanism at play. Mass quenching \citep[though e.g. AGN feedback or bar formation; see e.g.][and references therein]{Gavazzi2015} is expected to play an important role in the quenching of high-mass galaxies.  
For the rest, a possible quenching mechanism is `processing' in group environment before falling into the cluster. 

Not all galaxies arrive in the cluster from the field with largely unaffected gas reservoirs. In fact, simulations show that many of them are fed to the cluster from groups and in rare cases even from other clusters \citep{McGee2009}. Indeed, we find many groups currently being accreted into the cluster, accounting for some of the passive galaxies in the cluster outskirts. It is thus likely that at least some of the passive cluster galaxies were quenched in a previous group environment before entering the cluster (`pre-processing'). The morphologies of the galaxies and the X-ray properties of the groups suggest that in low mass groups, gas-rich interactions among galaxies are frequent. Such interactions typically result in an enhancement of star formation that may contribute to the consumption of gas, leading to a later quenching of the group galaxies. When groups grow and approach a mass $\sim10^{14}M\odot$, the interactions are less frequent. However, the intensity of ram-pressure increases, contributing to the quenching of the remaining star-forming galaxies. It is possible that additional mechanisms such as tidal interactions are also at play. 

Our results are consistent with the most recent multiwavelength study of galaxies in the substructures of the Virgo cluster \citep{Boselli2014}, that suggests that passive cluster galaxies have resulted from gravitational interactions in the infalling groups from which the cluster assembled. A previous BUDHIES analysis of the assembling cluster A2192\_1 also supports this scenario \citep[][]{Jaffe2012}. Simulations further show that mergers are key drivers of the morphological sequence in $\sim 10^{13} M\odot$ groups. The group galaxies in $<10^{13}M\odot$ haloes are expected to frequently experience mergers and galaxy-galaxy interactions, while galaxies in $10^{13}-10^{14}M\odot$ groups will be affected by groups tides  and RPS (e.g. strangulation). 
\newline

\textit{Post-processing: } In addition, we speculate that the presence of a large population of peculiar galaxies in the cluster centre  arise from the passage of groups through the cluster, that ``post-processes'' group and cluster galaxies by enhancing RPS and inducing galaxy mergers. Simulations \citep[][]{Vijayaraghavan2013} in fact show that the accretion of a group into a cluster can cause an enhancement of galaxy-galaxy collisions. Shock waves produced by the merger also enhance the ram-pressure intensity on group and cluster galaxies, and the increased local density leads to greater galactic tidal truncation. The combination of these phenomena can significantly transform (post-process) the cluster and group galaxies. This picture is consistent with recent WSRT observations of a merging cluster by \citet{Stroe2015}, that revealed gas rich  galaxies in the cluster with star formation activity triggered  by the passage of the shock. 
\newline

Our results highlight the entanglement of the different physical processes responsible for building up the large fraction of passive galaxies in clusters, and the need to account for the evolution of environment, imminent in a hierarchical Universe. 

\section*{Acknowledgements}

We thank the referee for providing constructive comments to the paper. 
YJ gratefully acknowledges Ricardo Demarco and FONDECYT grant no. 3130476 for their valuable support.
This work was co-funded under the Marie Curie Actions of the European Commission (FP7-COFUND). 
CPH was funded by CONICYT Anillo project ACT-1122. 
MMC gratefully acknowledges support from the Dunlap Institute. The Dunlap Institute is funded through an endowment established by the David Dunlap family and the University of Toronto. 
We are grateful for support from a Da Vinci Professorship at the Kapteyn Institute.
This work has been also supported by the Center for Galaxy Evolution Research funded by the NRF of Korea and Science Fellowship of POSCO TJ Park Foundation.
XF is supported by an NSF Astronomy and Astrophysics Postdoctoral Fellowship under award AST-1501342.
MY and RC acknowledge support from the NASA ADAP grant NNX14AF80G.
GPS acknowledges support from the Royal Society and the Science and Technology Facilities Council. 
NO is supported by a Grant-in-Aid from the Ministry of Education, 
Culture, Sports, Science, and Technology of Japan (26800097), the Funds for the Development of Human Resources in Science and Technology under MEXT, Japan, and Core Research for Energetic Universe in Hiroshima University (the MEXT program for promoting the enhancement of research universities, Japan). 
The Westerbork Synthesis Radio Telescope is operated by the ASTRON (Netherlands Institute for Radio Astronomy) with support from the Netherlands Foundation for Scientific Research (NWO). 
The SDSS is managed by the Astrophysical Research Consortium (ARC) for the Participating Institutions. The Participating Institutions are The University of Chicago, Fermilab, the Institute for Advanced Study, the Japan Participation Group, The Johns Hopkins University, Los Alamos National Laboratory, the Max-Planck-Institute for Astronomy (MPIA), the Max-Planck-Institute for Astrophysics (MPA), New Mexico State University, University of Pittsburgh, Princeton University, the United States Naval Observatory, and the University of Washington.
This work is based [in part] on observations made with the \textit{Spitzer Space Telescope}, which is operated by the Jet Propulsion Laboratory, California Institute of Technology under a contract with NASA. Support for this work was provided by NASA.
\textit{Herschel} is an ESA space observatory with science instruments provided by European-led Principal Investigator consortia and with important participation from NASA.
%



\appendix

\section{Improved spectroscopic completeness}
\label{sec:newz}

\begin{figure}
\centering
\includegraphics[trim=30 0 0 0,scale=0.52]{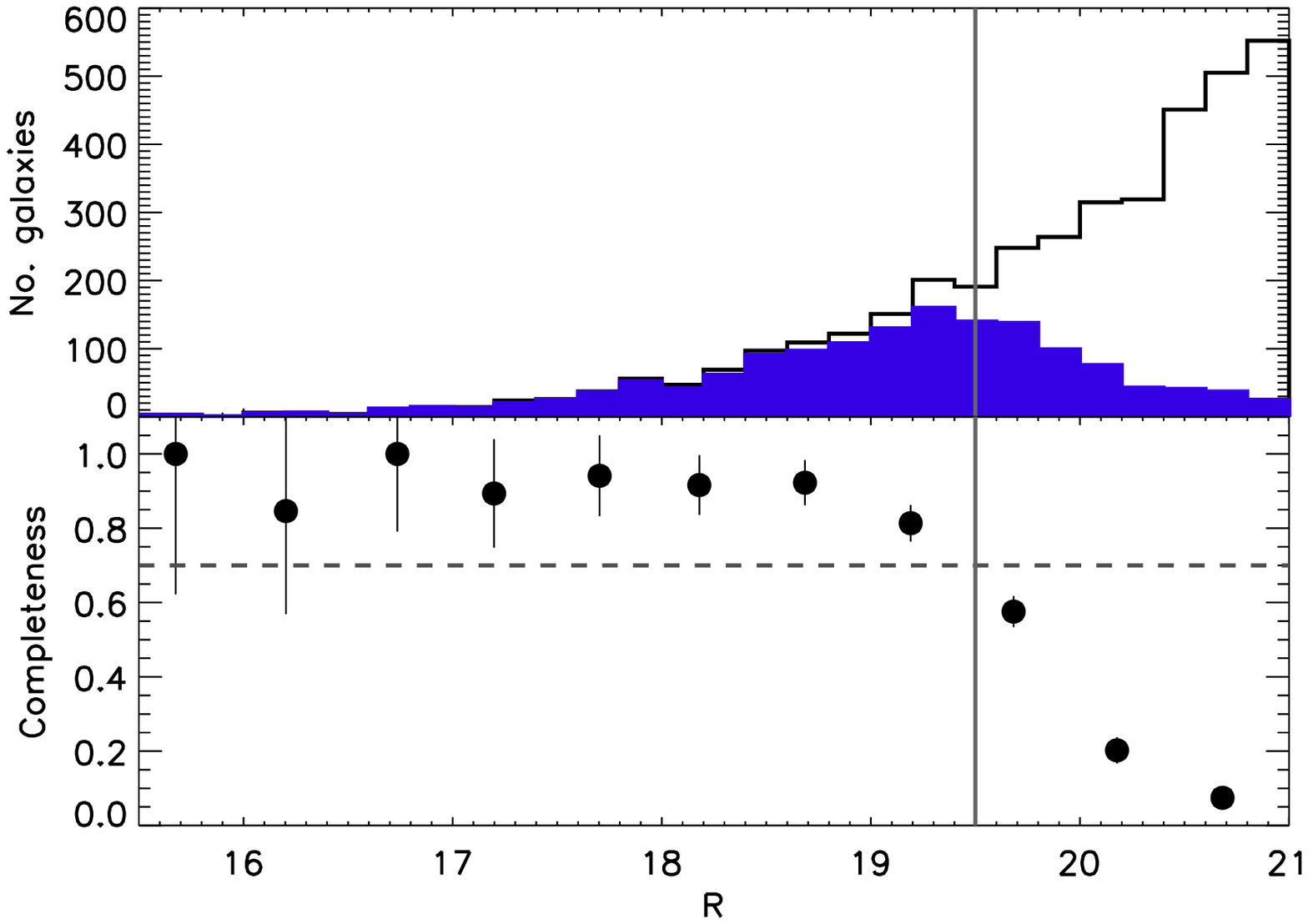}
\caption{ Spectroscopic completeness after inclusion of new redshifts. The open black histogram shows the R-band distribution of all the galaxies targeted for spectroscopy, while the blue histogram represents galaxies with a measured redshift. If we compare this plot with Figure 6 of \citet{Jaffe2013} it is clear that the inclusion of new redshifts (mainly from LoCuSS) has increased our spectroscopic completeness significantly ($>70$\%, dashed line) for galaxies with $R<19.5$ (solid line). 
 \label{completeness}}
\end{figure}

As mentioned in Section~\ref{sec:data}, we have included for this paper new redshifts in the field of Abell 963. This addition has improved our spectroscopic completeness significantly, as shown in Fig.~\ref{completeness}, where the distribution of the targeted galaxies for spectroscopy and those with a redshift are compared as a function of $R$-band magnitude. In particular, we are now complete ($>70$\%) down to $R=19.5$, as indicated by the lines in the bottom panel of the figure. 

\section{Results from the \HI\ stacks}
\label{sec:app_stacks}

This section presents \HI\ stacks that demonstrate the method and its reliability, and a summary of  all the \HI\ stacking results presented in this paper.

\begin{figure}
\centering
 \includegraphics[trim=0 0 0 0,scale=0.56]{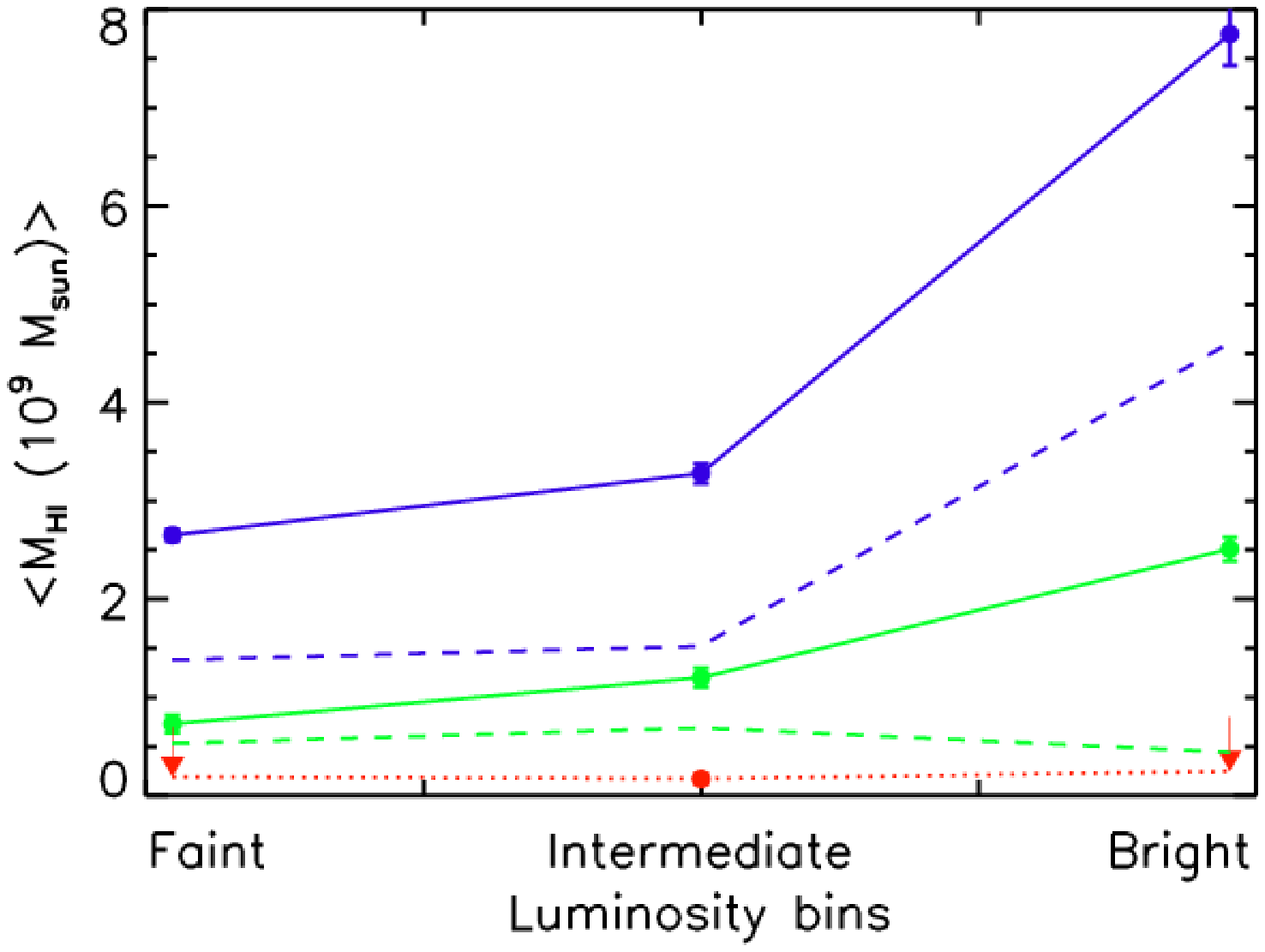}
\caption{ Results of the \HI\ stacking in different bins of colour and magnitude, presented in Fig.~\ref{HIstack_CMD}. Solid lines correspond to the stacks that consider all galaxies in each bin, and dashed lines to those that exclude direct \HI\ detections. When the signal was too low, and only an upper limit on the $M_{\HI\ }$ could be determined, arrows and dotted lines were used.   \label{HIstack_CMD_plot}}
\end{figure}

\begin{figure*}
\centering
 \includegraphics[trim=0 -14 0 0, scale=0.76]{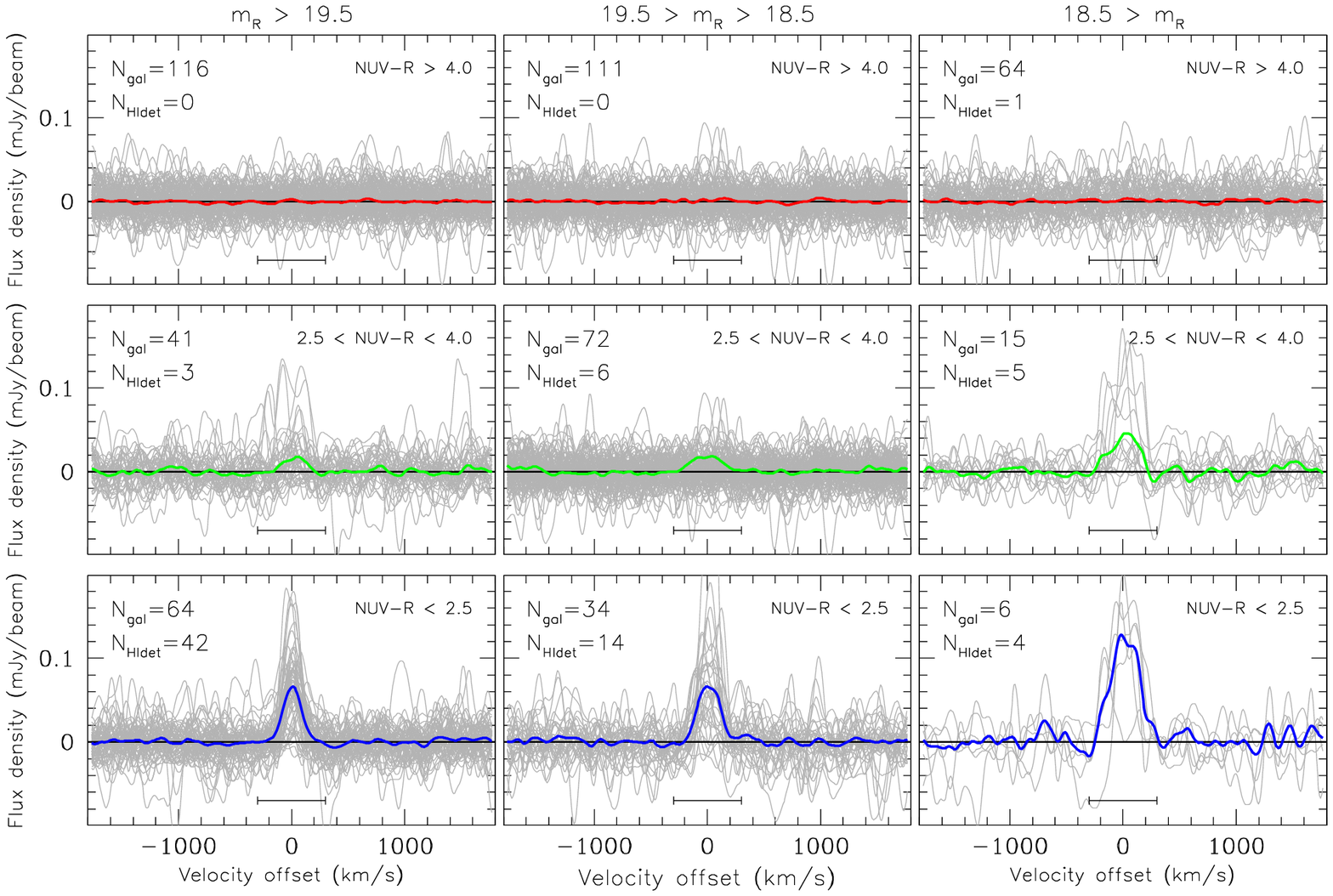}
  \includegraphics[trim=0 0 0 0, scale=0.76]{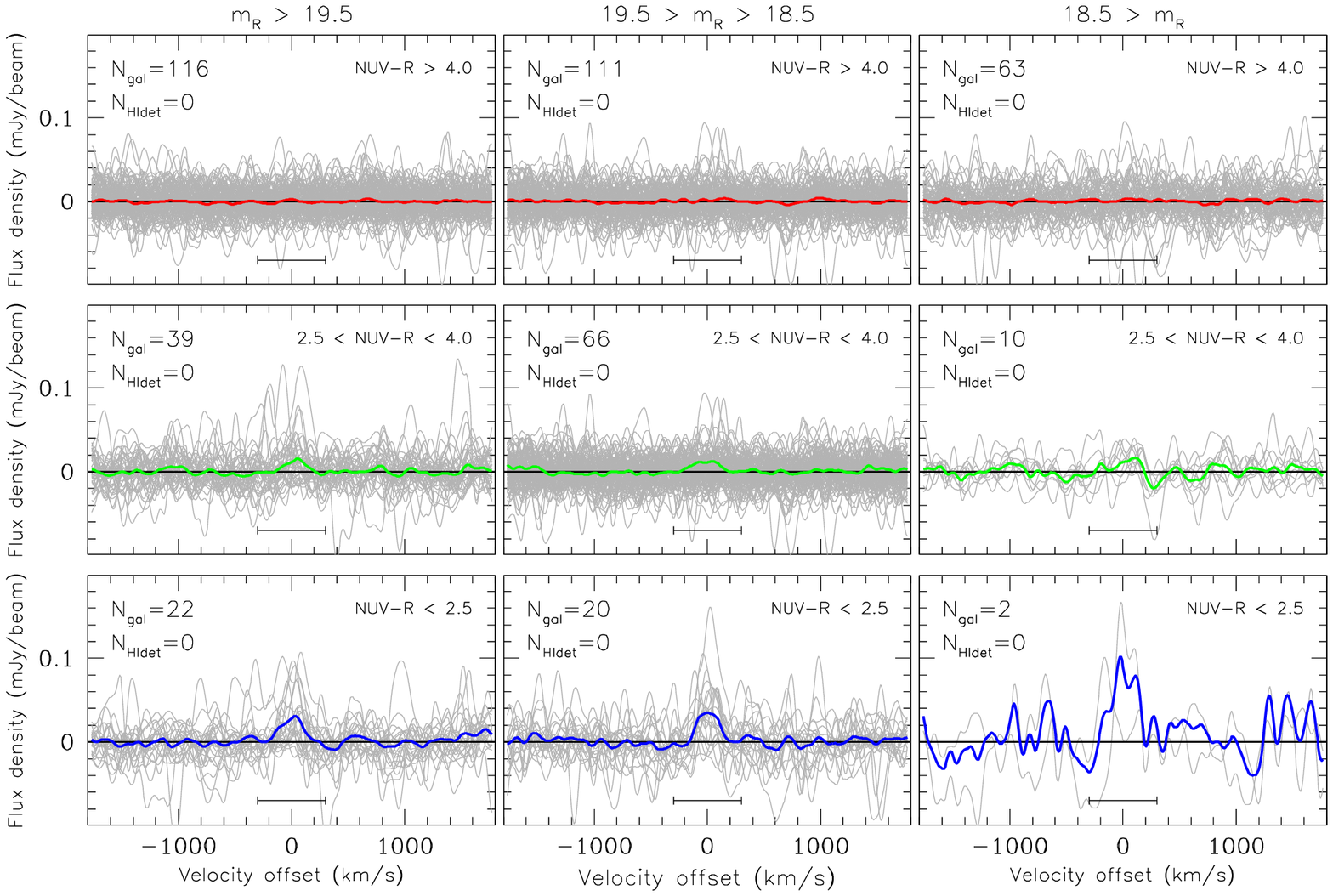}
\caption{\HI\ stacking in different bins of $NUV-R$ colour and $R$-band magnitude, as labelled (see Fig.~\ref{CMD} as reference). In each panel, the thin grey lines  correspond to individual galaxies while the thicker colour lines represent the stacked spectra. The number of galaxies that went into each stack ($N_{\rm gal}$) and number of direct \HI\ -detections ($N_{\HI\ {\rm det}}$) are indicated in the top-left corner of each panel. The top panel includes \HI\ -detected plus non-detected galaxies. The bottom panel shows the \HI\ stacks in the same bins, but excluding direct \HI\ -detections. 
 \label{HIstack_CMD}}
\end{figure*}

Fig.~\ref{HIstack_CMD} shows the \HI\ stacks in different bins of colour and magnitude, as labelled. The top panel includes all galaxies in each CMD bin while the bottom panel excludes direct \HI\ detections. It is noticeable how the 21~cm emission decreases with redder colours and fainter magnitudes. Moreover, the stacks that exclude direct \HI\ detections always present lower \HI\ fluxes, although the trend with luminosity and colour holds. This can be appreciated in Fig.~\ref{HIstack_CMD_plot}, where the average $M_{\HI\ }$ is plotted against the three luminosity bins considered, for each colour bin, for all galaxies (solid) and excluding direct \HI\ detections (dashed). 

Table~\ref{HIstack_table} lists the resulting \HI\ flux and mass for all the \HI\ stacks presented in this paper. 

\begin{table*}
\begin{tabular}{lccc}
\hline
\textbf{\HI\ stacking sample}					& No. galaxies	&  Flux		& $M_{\HI\ }$		\\
(see Section~\ref{subsec:HIstacks})				&		&  (mJy km ${\rm s}^{-1}$)	& $10^{9}M\odot$ 	\\
\hline
\textbf{Colour and magnitude:}					&		&			&			\\
Red/passive ($NUV-R>4$ or NUV undet.) and faint ($R>$19.5)	& 116		&  -0.25$\pm$0.31	& $<$0.19		\\
Red/passive ($NUV-R>4$ or NUV undet.) and int ($19.5>$R$>$18.5)	& 111		&  0.85$\pm$0.23	& 0.17$\pm$0.05		\\
Red/passive ($NUV-R>4$ or NUV undet.) and bright ($18.5>$R)	& 64		&  0.93$\pm$0.42	& $<$0.25			\\
Green (2.5$<$NUV-R$<$4) and faint ($R>$19.5)			& 41		&  3.62$\pm$0.44	& 0.73$\pm$0.09 	\\
Green (2.5$<$NUV-R$<$4) and int ($19.5>$R$>$18.5)		& 72		&  5.98$\pm$0.49	& 1.20$\pm$0.10 	\\
Green (2.5$<$NUV-R$<$4) and bright ($18.5>$R)			& 15		&  12.48$\pm$0.58	& 2.51$\pm$0.12		\\
Bluest (2.5$>$NUV-R) and faint ($R>$19.5)			& 64		&  13.19$\pm$0.31	& 2.65$\pm$0.06 	\\
Bluest (2.5$>$NUV-R) and int (9.5$>R>18.5$)			& 34		&  16.31$\pm$0.51	& 3.28$\pm$0.10		\\
Bluest (2.5$>$NUV-R) and bright int (18.5$>R$)			& 6		&  38.5$\pm$1.6		& 7.75$\pm$0.32		\\
\hline
\textbf{In/out the cluster ($R<$19.5):}				&		&			&			\\
Red/passive ($NUV-R>$4) and cluster core ($r<r_{500}$) 		& 131		& -0.04$\pm$0.21	& $<$0.13		\\ 
Red/passive ($NUV-R>$4) and cluster outskirts ($r>R_{200}$) 	& 126		& 1.69$\pm$0.34		& 0.34$\pm$0.07	\\
Blue ($NUV-R<4$) and cluster core ($r<r_{500}$) 			& 75		& 6.13$\pm$0.36		& 1.23$\pm$0.07	\\
Blue ($NUV-R<4$) and cluster outskirts ($r>R_{200}$) 		& 143		& 12.58$\pm$0.43	& 2.53$\pm$0.09	\\
\hline
\textbf{Groups by mass ($R<19.5$)}				&		&	 		&			\\
Red/passive in Groups A$+\delta$				& 5		& 5.6$\pm$1.4		& 1.13$\pm$0.28 	\\
Red/passive in Groups B$+\alpha$				& 20		& -2.85$\pm$0.62	& $<$0.37		\\
Red/Passive in Group C						& 27		& -0.31$\pm$0.68	& $<$0.41		\\
Blue in Groups A$+\delta$					& 11		& 14.69$\pm$0.72	& 2.96$\pm$0.14		\\
Blue and Groups B$+\alpha$					& 7		& 8.93$\pm$0.66 	& 1.80$\pm$0.13		\\
Blue and Group C							& 7		& 4.05$\pm$0.66		& 0.82$\pm$0.13		\\
\hline
\textbf{Phase-space regions	($R<19.5$):}			&		&			&			\\
Red/Passive in `recent infall' zone					& 81		& 2.37$\pm$0.46		& 0.48$\pm$0.09	\\
Red/Passive in `stripping' zone					& 28		& 0.08$\pm$0.64		& $<0.39$	\\
Red/Passive in `virialized' zone					& 28		& -0.90$\pm$0.36	& $<0.22$	\\
Blue in `recent infall' zone						& 85		& 13.50$\pm$0.44	& 2.72$\pm$0.09	\\
Blue in `stripping' zone						& 8		& 0.94$\pm$0.56		& $<0.34$	\\
Blue in `virialized' zone						& 7		& 1.06$\pm$0.79		& $<0.48$	\\
\hline
\end{tabular}
\caption{This table lists the results from the \HI\ stacks performed in this section and in Section~\ref{subsec:HIstacks}. Different lines correspond to different bins in colour, magnitude, regions in the cluster, groups, and phase-space zones as indicated in column 1. The second column displays the number of galaxies used in each stack, and columns 3 and 4 show the resulting \HI\ flux and estimated mass and errors.}
\label{HIstack_table}
\end{table*}

\end{document}